\documentclass[%
 reprint,
 amsmath,amssymb,
 aps,
 floatfix,
prb,
]{revtex4-2}

\usepackage{svg}
\usepackage{graphicx}
\usepackage{dcolumn}
\usepackage{bm}
\usepackage{physics}
\usepackage{dsfont}
\usepackage{hyperref, color, xcolor}
\usepackage{svg}
\usepackage{caption,subcaption}
\usepackage{soul}

\begin{document}

\preprint{APS/123-QED}

\title{Lowering the temperature of two-dimensional fermionic tensor networks with cluster expansions}

\author{Sander De Meyer$^{1,2}$}
\author{Atsushi Ueda$^1$}
\author{Yuchi He$^1$}
\author{Nick Bultinck$^1$}
\author{Jutho Haegeman$^1$}

\affiliation{$^1$Department of Physics and Astronomy, Ghent University, 
Krijgslaan 299, 9000 Gent, Belgium}

\affiliation{$^2$Center for Molecular Modeling, Ghent University, Technologiepark-Zwijnaarde 46, 9052 Zwijnaarde, Belgium}

\date{\today}

\begin{abstract}
Representing the time-evolution operator as a tensor network constitutes a key ingredient in several algorithms for studying quantum lattice systems at finite temperature or in a non-equilibrium setting. For a Hamiltonian composed of strictly short-ranged interactions, the Suzuki-Trotter decomposition is the main technique for obtaining such a representation. In [B.~Vanhecke, L.~Vanderstraeten and F.~Verstraete, Physical Review A, L020402 (2021)], an alternative strategy --the cluster expansion-- was introduced. This approach naturally preserves internal and lattice symmetries and can more easily be extended to higher-order representations or longer-ranged interactions. We extend the cluster expansion to two-dimensional fermionic systems, and employ it to construct projected entangled-pair operator (PEPO) approximations of Gibbs states. We also discuss and benchmark different truncation schemes for multiplying layers of PEPOs together. Applying the resulting framework to a two-dimensional spinless fermion model with attractive interactions, we resolve a clear phase boundary at finite temperature.
\end{abstract}

\maketitle

\section{\label{sec:introduction}Introduction}
Fermionic many-body systems in two dimensions can exhibit rich physics. Strong correlations and the possible presence of a Fermi surface can yield behavior far beyond the reach of perturbation theory, including high-$T_c$ superconductivity. Many such systems can be modeled as fermions hopping on a lattice and interacting locally. Historically, quantum Monte Carlo (QMC) methods have been remarkably successful in this setting. However, in doped regimes, they often encounter the notorious sign problem, which quickly renders simulations infeasible. Tensor networks (TNs) \cite{TN1, TN2, TN3} have emerged as an alternative way to overcome this problem. In one dimension, TNs, often referred to as matrix product states (MPS), admit the powerful density matrix renormalization group (DMRG) algorithm and have achieved striking success in capturing strong correlations. The real challenge, however, lies in the application to genuinely two-dimensional quantum systems.

While many studies have used snake-like MPSs to investigate two-dimensional strips or cylinders, recent results have also shown the potential of using projected entangled-pair states (PEPS) to investigate two-dimensional systems directly in the thermodynamic limit, including at finite temperatures \cite{PEPS_original, Finite_T_before_Czarnik1, Finite_T_before_Czarnik2, Czarnik_1, Czarnik_2, Czarnik_3}. These algorithms have shown their power to determine the critical temperature and exponents of both bosonic models, such as the Ising model \cite{Czarnik_1, Czarnik_3, Czarnik_4, FCLS_Corboz_Czarnik}, a spinless fermion model \cite{Czarnik_2}, and the Hubbard model \cite{PEPS_Finite_T_Hubbard1, Variational_PEPS_Finite_T2}. In regimes accessible to QMC, these TN approaches obtain comparable results, but they can also be applied in regimes where QMC exhibits sign problems, notably when doped away from half-filling.

The finite-temperature partition function $Z(\beta) = \Tr{\exp(-\beta H)}$ of a two-dimensional quantum system with a Hamiltonian $H$ --assumed to be composed of local interactions-- can be represented as a three-dimensional TN. Hereto, firstly, the imaginary-time direction is discretized into steps $\Delta \beta$, and the evolution operator $\exp(-\Delta \beta H)$ is approximated as a projected entangled-pair operator (PEPO) up to some accuracy in the step size $\Delta \beta$. These PEPO layers are stacked together, and the resulting three-dimensional network needs to be contracted. This, secondly, requires further approximations, for which a range of alternative algorithms have been proposed in the literature. In the thermodynamic limit, it is most natural to first contract the network along the finite temporal direction, \textit{i.e.} to multiply the different PEPO layers together, into a single PEPO that approximates the Gibbs state $\rho(\beta) = Z(\beta)^{-1}\exp(-\beta H)$. The first step of the algorithm, obtaining a PEPO representation of $\exp(-\Delta \beta H)$, is typically performed using a first or second-order Suzuki-Trotter (ST) decomposition~\cite{Suzuki1976, Trotter1959}. The maximal time step $\Delta \beta$ must thus be chosen sufficiently small to keep the Trotter discretization error under control. As a consequence, many PEPO layers need to be contracted to reach the low temperature regime, so that both Trotter errors and truncation errors can accumulate. To overcome this issue, an alternative expansion of the time-evolution operator has recently been proposed, that maintains extensivity by partitioning different contributions in terms of a maximal size of connected clusters~\cite{ClusterExpansion1, ClusterExpansion2, TTNR}. This ``cluster expansion'' (CE) approach exhibits a more favorable accuracy in terms of the imaginary-time step~\cite{ClusterExpansion1, ClusterExpansion2, TTNR} and naturally preserves internal and lattice symmetries of the Hamiltonian, including full translation invariance. In the present work, we extend the formalism of cluster expansions to fermionic models and show that we can reach significantly lower temperatures accurately. Hereto, we also benchmark three different temporal contraction or truncation strategies.

The resulting methodology is then used to investigate the phase diagram of spinless fermions hopping on the square lattice and interacting via a nearest-neighbor density interaction at half-filling. This model exhibits phase separation, stripe order, and nematic order and can be used to describe several classes of superconductors~\cite{SF_mean_field, SF_SC_Girvin_1978, SF_SC1, SF_SC2, SF_SC3, SF_SC4_1, SF_SC4_2}. Previous studies have used mean-field methods~\cite{SF_mean_field}, QMC~\cite{SF_QMC}, or tensor networks~\cite{Czarnik_2}. While the sign problem in QMC can be remedied for repulsive interactions, this is not the case for attractive interactions~\cite{QMC_sign_problem}, which is the regime of study in this paper. Mean-field methods are able to calculate the different phase transitions and can tell us which phases are present in which parameter regimes, but also suffer from limited accuracy. The most accurate method of determining the phase diagram is thus via tensor networks. 
While the earlier TN study in Ref.~\onlinecite{Czarnik_2} has been crucial in the development of finite-temperature tensor network calculations, they have only studied the thermal phase transition in this particular model for a very large interaction strength, where the transition temperature is relatively high, so that it could be obtained using the ST decomposition without noticeable issues. This paper aims to improve upon these results using the cluster expansions approach in order to scan the phase diagram, and thus the transition temperatures, for a whole range of interaction strengths in the weak and intermediate regime.

The paper is organized as follows. We start with a short overview of the different building blocks of our methodology in the next section. This includes a short discussion on fermionic tensor networks in Subsection~\ref{sec:fermionicTN}, an introduction to the cluster expansions as a means of obtaining a controlled approximation of the evolution operator in Subsection~\ref{sec:methodology}, and an explanation about the different truncation schemes for the imaginary time evolution or temporal contraction in Subsection~\ref{sec:truncation_schemes}, where we end with the proposal of a new variational scheme. In the results section we will show several benchmarks. In section \ref{sec:Comparison_CE_Trotter_SU}, we show that the cluster expansions lead to a more accurate representation of the density operator of a spin model, allowing us to use larger steps in the imaginary-time evolution. Subsection \ref{sec:SF} introduces a spinless fermion model with attractive interactions that is used for the final simulations. In section \ref{sec:comparison_truncation_schemes}, the different truncation schemes are compared. In section \ref{sec:phase_diagram}, we explain how to perform the scaling analysis and present the final phase diagram of the spinless fermion model. Finally, Section~\ref{sec:conclusion} contains our conclusions and an outlook towards future applications and extensions of the proposed workflow.

\section{Methodology}

\subsection{\label{sec:fermionicTN}Fermionic tensor networks}

The fermionic nature of states and operators can be dealt with in tensor networks by using the framework of super vector spaces \cite{Fermion_Nick}. A super vector space $V$ has an orthogonal direct sum decomposition
\begin{equation}
    V = V^0 \oplus V^1 \; ,
\end{equation}
with vectors $\ket{v}$ having support in only one of these orthogonal sectors being called homogeneous. A homogeneous vector $\ket{v}$ has parity $|v|$ and is called even (odd) if it has parity 0 (1) and is an element of $V^0$ ($V^1$). The (graded) tensor product of two homogeneous vectors $\ket{u}$ and $\ket{v}$ is a homogeneous vector $\ket{u} \otimes_g \ket{v}$ with parity $|u| + |v| \mod 2$, corresponding to the fusion rules of $\mathbb{Z}_2$. The link between super vector spaces and fermions becomes clear when considering the isomorphism
\begin{eqnarray}
&\mathcal{F}: V_1 \otimes_g V_2 \xrightarrow{} V_2 \otimes_g V_1 :  \nonumber\\
&\ket{u}_1 \otimes_g \ket{v}_2 \xrightarrow{} (-1)^{|u| |v|} \ket{v}_2 \otimes_g \ket{u}_1 \; ,
\label{eq:fermionic_reordering}
\end{eqnarray}
which is called fermionic reordering. This isomorphism exactly encodes the anticommutation relations necessary for describing fermionic quantum many-body systems. When implemented as an internal symmetry, tensor contractions can be performed if the permutation of internal legs follows the reordering procedure given by $\mathcal{F}$. While the fermionic nature of the tensor network can be treated as a black box in the open-source software package TensorKit.jl~\cite{TensorKit} for most contractions, there are some subtleties where the composition of operators, or the computation of inner products or traces, do not naively map to the fermionic contractions of the corresponding tensor networks, but require some compensation of sign factors by inserting additional fermionic parity tensors in the network. We refer to Ref.~\onlinecite{Mortier_2025} for further details.

\subsection{\label{sec:methodology}Cluster expansions}
Equilibrium physics at finite temperature is encapsulated in the Gibbs state and the corresponding partition function
\begin{equation}
    \rho(\beta) = \frac{1}{Z(\beta)} e^{-\beta H}, \; \; \; Z(\beta) = \text{Tr}\left(e^{-\beta H}\right) \; ,
\end{equation}
where $H$ and $\beta$ are the Hamiltonian and inverse temperature, respectively. Unfortunately, exactly computing the exponential of an extensive operator consisting of non-commuting terms is unfeasible.

All approaches for computing the exponential start by ``discretizing the temporal direction'', \textit{i.e.} by rewriting the exponential as a composition $e^{-\beta H} = [e^{-\Delta \beta H} ]^N$ where $\Delta \beta = \beta/N$ is a small (imaginary) time step, and then further approximating $e^{-\Delta \beta H}$. While it might seem natural to approximate $e^{-\Delta \beta H}$ by its Taylor expansion up to some order, this does not yield an approximation that is extensive, as the different orders have a completely different scaling as function of system size, and would end up having vanishing contributions in the thermodynamic limit. Instead, the most common approach, at least when $H$ consists of a sum of strictly local terms, is to employ the ST decomposition, which gives rise to a product of 2-site gates (for nearest-neighbor terms). In the context of PEPS, the ST decomposition is the first step behind a number of algorithms for approaching both ground states \cite{SU_original, SU_Banuls, SU_Jahromi_2019, FFU} and finite-temperature states \cite{Czarnik_1, Czarnik_2, Czarnik_3, Czarnik_4, Variational_PEPS_Finite_T2, FCLS_Corboz_Czarnik}. These algorithms then differ in the way the subsequent gates are composed or applied. They can be composed to contruct the PEPO representing $e^{-\Delta \beta H}$, which we will call the `ST PEPO', or directly applied to a bond of the thermal density operator in algorithms like Simple Update (SU) or (Fast) Full Update (FU/FFU) \cite{SU_original, SU_Banuls, SU_Jahromi_2019, FFU}. Despite its many successes, some drawbacks of the ST decomposition include that it often requires breaking either spatial or internal symmetries of the system, and that the complexity of higher-order decompositions increases quickly. Hence, one often restricts to first or second-order ST decompositions, which then implies being limited to small (imaginary) time steps. The breaking of translational invariance furthermore implies an extra overhead for the ST decomposition, as the resulting cost scales linearly in the size of the unit cell for both the time evolution algorithm as the environment calculations.

Instead, we here use an expansion of the exponential that is organized in terms of the maximal size of connected clusters, and which can naturally be represented as TN operator, or thus a PEPO in two dimensions. This idea was pioneered in Ref.~\onlinecite{ClusterExpansion1} and further investigated in Ref.~\onlinecite{ClusterExpansion2}. The general idea of using expansions where the maximal size of connected clusters is the expansion parameter has a rich history in statistical physics, and closely related expansions were used in mathematical proofs regarding the polynomial scaling of tensor network bond dimension in representing ground states and thermal states of locally interacting systems \cite{Hastings,Kliesch,Molnar} or in the context of stochastic series expansions \cite{Sandvik}. For tensor networks, this approach is motivated by the observation that, given a tensor network representation of an operator on a local patch of the lattice, it is straightforward to create an extensive and translation-invariant superposition over products of this operator acting on non-overlapping patches. This construction, remeniscent of a finite-state machine, is easiest to visualize in 1D. For example, if we are considering a local operator on a 2-site patch that can be written as $O = L R$, a translation-invariant extensive operator containing higher-order powers of this operator is obtained by as a tensor network operator constructed from the local tensor
\begin{equation}
    \begin{bmatrix}
    \mathds{1} & L\\
    R & 0
\end{bmatrix}
\end{equation}
on every site of the 1D lattice.

To end up with the construction of the cluster expansion, we start from the Taylor expansion of the exponential, \textit{i.e.}
\begin{equation}
    \exp(-\Delta \beta H) = \sum_{k=0}^{+\infty} \frac{(-\Delta \beta)^k}{k!} H^k \; ,
\label{eq:TaylorExpansion}
\end{equation}
with $H = \sum_\ell h_\ell$ a Hamiltonian consisting of local terms. Expanding $H^k$ will result in a sum of products of $k$ Hamiltonian operators $h_\ell$. Some of the factors of each product will have overlapping support, but every term of $H^k$ can be reinterpreted as a product of some nontrivial operators acting on one or more disjoint clusters. We can then organize all the terms of equation \eqref{eq:TaylorExpansion} by the size of the largest cluster that is contained in this disjoint cluster product, leading to an expansion 
\begin{equation}
\exp(-\Delta \beta H) = \sum_{p = 0}^\infty \mathcal{T}_p \; .
\end{equation}
Here, $\mathcal{T}_p$ contains all terms where the largest disjoint cluster is of size $p$, with in particular, $\mathcal{T}_0 = \mathds{1}$. In fact, for the larger cluster sizes, we can further partition the expansion more finely so as to also reflect the shape of the largest cluster.

We now consider the structure of this reorganized expansion when restricting to a maximum cluster size $P$
\begin{equation}
\exp(-\Delta \beta H) = \sum_{p = 0}^{P} \mathcal{T}_p + \mathcal{O}(\Delta \beta^{n+1}).
\end{equation}
Firstly, the summation in the right-hand side contains all terms from the original Taylor expansion up to order $n$, if $P$ is the largest cluster size that can be covered by an overlapping product of $n$ Hamiltonian terms. For a Hamiltonian with nearest-neighbor terms only, this leads to $n = P-1$. Secondly, the summation of all the terms up to some maximal cluster of size $P$ can exactly be recognized as the translation-invariant and size-extensive superposition of the operator that corresponds to $\exp(-\Delta \beta \left.H\right|_P)$, where $\left.H\right|_P$ is the restriction of the Hamiltonian onto one such cluster. Put differently, $\left.H\right|_P$ contains all terms of the Hamiltonian that are confined to a single cluster of size $P$. 

To make this more explicit, let us now illustrate the corresponding tensor network construction for the case of a nearest-neighbor Hamiltonian, i.e.
\begin{equation}
    H = \sum_{\left< i,j \right>} h_{i,j}^{(2)} \; .
\label{eq:nn_hamiltonian}
\end{equation}
with clusters going up to size $P=3$. This is the case that will be used throughout this work. Hereto, we start by defining local tensors for the smallest cluster of size $p=2$
\begin{equation}
\begin{split}
    &\exp \left(\vcenter{\hbox{\includegraphics[page=8, scale=0.8]{Typst/ClusterExpansions.pdf}}} \right) - \vcenter{\hbox{\includegraphics[page=11, scale=0.8]{Typst/ClusterExpansions.pdf}}} \\ 
    &= \vcenter{\hbox{\includegraphics[page=9, scale=0.8]{Typst/ClusterExpansions.pdf}}} = \vcenter{\hbox{\includegraphics[page=10, scale=0.8]{Typst/ClusterExpansions.pdf}}}\; . \\
\end{split}
\label{eq:Construction_CE_H2}
\end{equation}
and then additional tensors to capture the novel contributions originating on clusters of size $p=3$
\begin{equation}
\begin{split}
    &\exp \left(\vcenter{\hbox{\includegraphics[page=2, scale=0.8]{Typst/ClusterExpansions.pdf}}} + \vcenter{\hbox{\includegraphics[page=3,scale=0.8]{Typst/ClusterExpansions.pdf}}} \right) \; \\
    &- \left(\vcenter{\hbox{\includegraphics[page=5,scale=0.8]{Typst/ClusterExpansions.pdf}}} + \vcenter{\hbox{\includegraphics[page=6,scale=0.8]{Typst/ClusterExpansions.pdf}}} + \vcenter{\hbox{\includegraphics[page=7,scale=0.8]{Typst/ClusterExpansions.pdf}}} \right)\\
    &= \vcenter{\hbox{\includegraphics[page=1, scale=0.8]{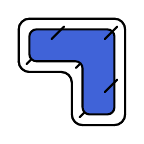}}} = \vcenter{\hbox{\includegraphics[page=4, scale=0.8]{Typst/ClusterExpansions.pdf}}} \; .
\end{split}
\label{eq:Construction_CE_H3}
\end{equation}
These local tensors can be combined into a single PEPO where all clusters up to size $P$ are taken into account exactly. Furthermore, in many cases the problem for the next cluster size can be parameterized by recycling the tensors from the smaller cluster and only adding a single new cluster, which is then determined by a linear equation. In the example of equations \eqref{eq:Construction_CE_H2} and \eqref{eq:Construction_CE_H3}, this means that we can choose $C = A$ and $E = B$, and find the new tensor $D$ using a linear solver. As a consequence, we need fewer local tensors to represent the PEPO, which will thus have a lower total bond dimension. The bond dimension of the $P=3$ cluster expansion will be equal to $D=5$ for both the quantum Ising model and the model of spinless fermions considered in this paper, and the resulting expansion is accurate up to second-order ($n=2$) \footnote{
For the quantum Ising model, the special structure of the terms allows us to build a second-order symmetric Trotter decomposition as a translation-invariant PEPO with bond dimension $D=2$. While the $D=5$ cluster expansion PEPO has the same scaling of the error, it is expected to contain several higher-order contributions that are not present in the Trotter PEPO, so that the factor in front of the order-$(\Delta \beta)^3$ error would be smaller. For a generic nearest-neighbor Hamiltonian, even the first-order Trotter expansion would result in a PEPO with $D=4$ with a 2-site unit cell (broken translation and rotation invariance), whereas a second-order Trotter expansion would exhibit a PEPO representation with $D=16$ on most of the bonds.
}.
Since the extra cost of calculating the second-order ($P=3$) contribution is negligible and doesn't increase the PEPO bond dimension, we will always use the resulting expansion in this paper. Going to higher orders can be useful if a higher accuracy is desired for a given time step, but it will increase the computational cost of the time evolution via the bond dimension of the resulting PEPO. Further details of this construction can be found in appendix \ref{ap:CE} and references \cite{ClusterExpansion1, ClusterExpansion2}. This construction can be generalized to longer-range Hamiltonians, but it requires contributions from larger clusters to maintain the same level of accuracy. There exist alternative constructions in one and higher dimensions, which furthermore works for long-range interactions, based on resumming the disconnected terms in the Taylor expansion up to a certain order \cite{Pollman2015, Maarten_Taylor, dunham2025, Sotnikov}. While for nearest-neighbour Hamiltonians, the considered terms are a proper subset of the terms considered in a cluster expansion of the same order, the bond dimension of the resulting MPO or PEPO scales better for long-range interactions. More details can be found in \cite{Maarten_Taylor} for a generic construction in 1D and \cite{dunham2025} for a second-order construction in 2D.

Having constructed the PEPO approximation to $\exp(-\Delta \beta H)$, we will now start multiplying these layers to reach a final (inverse) temperature $\beta$. Combining $N = \beta / \Delta \beta$ layers would result in a final bond dimension $D^N$, which is of course unfeasible. Intermediate truncation steps are required, using strategies that will be discussed in the next subsection. For the composition process itself, two main strategies can be devised. We can update one PEPO of bond dimension $D_1$ approximating $e^{-\beta_k H}$ by composing it with our elementary PEPO approximation of $e^{-\Delta \beta H}$ with bond dimension $D_2=5$ to result in a PEPO approximating $e^{-\beta_{k+1} H}$, where $\beta_k = k \Delta \beta$. Alternatively, we successively double the inverse temperature, \textit{i.e.}\ we reach $\tilde{\beta}_{k+1}$ by squaring a PEPO representation of $e^{-\tilde{\beta}_k H}$ where $\tilde{\beta}_k = 2^k \Delta \beta$. The latter approach was applied in the context of matrix product states under the name of ``exponential tensor renormalization group'' \cite{Chen}, and allows to reach low temperatures more quickly. However, as we now have to multiply two equal layers with some bond dimension $D_1$ and then truncate the result, the computational cost will be higher, thus limiting the maximal $D_1$ to represent the final finite-temperature state. Furthermore, because of the larger truncation required, the accumulated errors might increase more quickly, although this may be offset by the fewer number of multiplications required. Nonetheless, as our main goal is scanning the finite-temperature phase diagram, we will take the former approach in this work, resulting in a finer resolution along the temperature axis.

\subsection{\label{sec:truncation_schemes}Truncation schemes}
We now turn to the problem of truncating the bond dimension when composing several layers of PEPOs. Henceforth, we denote the density operator represented as a PEPO with local tensor $A$ as a state $\ket{A}$, by invoking the Choi–Jamio\l{}kowski isomorphism. For convenience, we will consider unit cells consisting of a single tensor. The proposed methods can be easily extended to larger unit cells. When composing a PEPO $\ket{A}$ with bond dimension $D_1$ with a PEPO $\ket{\delta A}$ with bond dimension $D_2$, the exact application of the resulting PEPO $\ket{A^\prime}$ is given by the product $D_1 D_2$. As repeating this procedure would quickly become infeasible, the new PEPO has to be truncated to a PEPO $\ket{B}$ with a smaller bond dimension such that the state is minimally disturbed. 

The truncation strategy is of crucial importance to both the accuracy and the computational complexity of the resulting algorithm for constructing thermal states, and similarly appears in the context of real-time evolution. One of the biggest questions in this respect is the extent to which the environment of the tensors influences this truncation, and how to strike the balance between precision and computational cost. A great variety of different strategies have been proposed. Initially, the truncation was performed using explicit isometries that map the enlarged Hilbert space to a truncated space with a lower bond dimension. Using horizontal and vertical isometries, called $T_h$ and $T_v$ respectively, satisfying
\begin{eqnarray}
    T_h: \mathcal{H}_h(A) \otimes \mathcal{H}_h(\delta A) \leftarrow \mathcal{H}_h(B) \; , \nonumber\\
    T_v: \mathcal{H}_v(A) \otimes \mathcal{H}_v(\delta A) \leftarrow \mathcal{H}_v(B) \; ,
\label{eq:isometries_definition}
\end{eqnarray}
with $\mathcal{H}_h$ ($\mathcal{H}_v$) the Hilbert space on the horizontal (vertical) bond of the tensor, the new tensor $B$ is given by
\begin{equation}
    \vcenter{\hbox{\includegraphics[page=3]{Typst/PEPO_Applications.pdf}}} = \vcenter{\hbox{\includegraphics[page=2]{Typst/PEPO_Applications.pdf}}}\label{eq:defBviaisometries}
\end{equation}
The dimension of $\mathcal{H}_v(B)$ is a measure of the accuracy and computational complexity of the simulation. In the earliest works, these isometries were constructed from the singular value decomposition (SVD) in the full environment of the untruncated tensor $A'$ \cite{Czarnik_1,Czarnik_2}, though this was later replaced by a self-consistent environment of the truncated tensors \cite{Czarnik_3}. Next, by representing the successive truncations as a tree tensor network, a direct variational maximization of the full partition function known as ``Variational Tensor Network Renormalization" was investigated \cite{Czarnik_4,Czarnik_5,Czarnik_6,Variational_PEPS_Finite_T2}. In all of these, the tensor environments are typically computed using the Corner Transfer Matrix Renormalization Group (CTMRG) algorithm \cite{CTMRG1, CTMRG2, CTMRG3}. However,  in some cases, the completely local ``simple update'' or ``belief propagation'' environments can be used \cite{Kshetrimayum_thermal,Kshetrimayum_time,Hubig_time,beliefpropagation,tindall}.

Instead of constructing the new tensor $B$ by explicitly applying isometries to the bigger tensor $A'$, it can also be obtained via the direct variational optimization (maximization) of the overlap or fidelity $f$ between the untruncated and truncated PEPOs, which is given by
\begin{equation}
    f^2 = \left| \braket{B}{A'} \right|^2 / (\braket{B}{B} \braket{A'}{A'}) \; .\label{eq:costfunction_full}
\end{equation}
This variational approach has become more standard recently, especially in the context of real-time evolution. Once again, different strategies are possible to evaluate or estimate the cost function and the gradient or search direction. As before, the main variability is in the accuracy of the environment. The first approaches, based on the full CMRG environment are known as ``exact environment full update'' \cite{Czarnik_8,Czarnik_9,Czarnik_10}. However, promising results have also been obtained with local approximations of the environment, which is known as the ``neighborhood tensor update''\cite{Dziarmaga_NTU,PEPS_Finite_T_Hubbard1} and is closely related to earlier ``cluster update'' proposals in the context of finite-sized PEPS ground state simulations \cite{Wang_Cluster,SU_Banuls}.

We now discuss three different strategies in more detail, with increasing degree of complexity. In the next section, we compare these three different strategies in their ability to find the optimally truncated tensor $B$, as expressed via the fidelity $\left|\braket{B}{A'}\right|$, as function of the target bond dimension.

\subsubsection{Local truncation}
The first strategy we discuss constructs isometries explicitly and then defines $B$ via Eq.~\eqref{eq:defBviaisometries}, and was recently used to study thermal phase transitions \cite{TTNR,Naravane}. The construction of the isometries is inspired by the Boundary Tensor Renormalization Group (BTRG), as detailed in Ref.~\onlinecite{BTRG}. The aim of this method is to minimize
\begin{equation}
    \epsilon = \left| \vcenter{\hbox{\includegraphics[page=5,scale=0.7]{Typst/PEPO_Applications.pdf}}} - \vcenter{\hbox{\includegraphics[page=6,scale=0.7]{Typst/PEPO_Applications.pdf}}} \right|^2 \; .
\end{equation}
The isometries are obtained from a SVD, with a preprocessing step involving QR and LQ decompositions to lower the computational cost. The dominating costs of this approach have a complexity of $d^3 D_1^5$ and $d^3 D_2^5$ for the QR/LQ decompositions and $d^2 D_1^3 D_2^3$ for the SVD decomposition, with $d$ the physical dimension and $D_1$ ($D_2$) the bond dimension of the top (bottom) layer. This approach optimizes the difference between the local tensors without taking any environment into account. As such, the resulting truncated tensor $B$ is not expected to maximize the overlap.

\subsubsection{Global truncation}
Taking the environment into account can yield more accurate results, but will also be computationally more demanding for large environment bond dimensions. We here generalize the iterative approach of Ref.~\cite{Czarnik_3}, where it was applied in conjunction with the ST decomposition. In particular, an initial guess of the isometries $T_h$ and $T_v$ define an initial $B$. We then calculate the CTMRG environment of the corresponding `trial state' $\ket{B}$
\begin{equation} \vcenter{\hbox{\includegraphics[page=3,scale=1]{Typst/VOMPS_PEPO.pdf}}} = \vcenter{\hbox{\includegraphics[page=6,scale=1]{Typst/VOMPS_PEPO.pdf}}} \; ,
\label{eq:double_layer_global}
\end{equation}
where the environment of the square-lattice PEPO is denoted as a box around the local tensor in the right-hand side, and the dots in the left hand side denote the infinite repetition of the tensors on the square lattice. This environment is used to construct the tensor
\begin{equation} \vcenter{\hbox{\includegraphics[page=10,scale=1]{Typst/VOMPS_PEPO.pdf}}} = \vcenter{\hbox{\includegraphics[page=9,scale=1]{Typst/VOMPS_PEPO.pdf}}} \; ,
\end{equation}
where the green tensors $A^\prime$ represent the exact application of $\delta A$ on $A$, but where the current isometries are applied along those directions that are contacted with the environment \footnote{The isometries are thus applied to all directions but the right one for the left tensor and vice versa.}. New isometries $T_h$ and $T_v$ are computed by performing an SVD on the tensor $E$, which are then used in the next iteration. This procedure is iterated a number of times until some convergence criterion or a maximal number of iterations is reached. The computational cost of this approach is dominated by the repeated CTMRG calculations, which have a complexity that scales as $D_1^6 \chi^3$, which $\chi$ the bond dimension of the CTMRG environment.

\subsubsection{Variational truncation}
As a final strategy, we can directly maximize the fidelity in Eq.~\eqref{eq:costfunction_full} as a nonlinear optimization problem. Since the norm of $A'$ is fixed, we can alternatively minimize
\begin{equation}
    \epsilon^2 = -\frac{|\braket{B}{A'}|^2}{\braket{B}{B}} \; ,
\label{eq:costfunction_simple}
\end{equation}
such that the calculation of a 4-layer environment is avoided. The expression for $\epsilon$ above can be visualised as
\begin{equation}  \vcenter{\hbox{\includegraphics[page=5,width=0.3\linewidth]{Typst/VOMPS_PEPO.pdf}}} \times \left( \vcenter{\hbox{\includegraphics[page=6,width=0.3\linewidth]{Typst/VOMPS_PEPO.pdf}}} \right)^{-\frac{1}{2}} \; .
\label{eq:costfunction_simple_visualization}
\end{equation}
As before, the evaluation of these infinite tensor networks can again be done using CTMRG.

The gradient of this cost function is, up to an overall constant, equivalent to the gradient of the cost function $\tilde{\epsilon}^2 = \norm{\ket{B} -\ket{A'}}^2$. The exact gradient of the cost function, before invoking the CTMRG calculation, is obtained by removing a single tensor $\overline{B}$ from the infinite networks. If we again approximate the rest of the network with the CTMRG environments, the gradient is given by
\begin{equation} 
\vcenter{\hbox{\includegraphics[page=7,width=0.3\linewidth]{Typst/VOMPS_PEPO.pdf}}} - \vcenter{\hbox{\includegraphics[page=8,width=0.3\linewidth]{Typst/VOMPS_PEPO.pdf}}}\label{eq:fidelitygradient} \; .
\end{equation}
Temporarily ignoring the $B$ dependence in the environments, the condition of vanishing gradient can be used to turn Eq.~\eqref{eq:fidelitygradient} into a linear system $G(B) - Z=0$ for $B$, where $G(B)$ represents the action of the blue CTMRG environment acting on $B$, whereas $Z$ represents the full red diagram. The solution of this system, $B = G^{-1}(Z)$, can be used as a new iterate, or as a search direction along which to find a new $B^\prime(\alpha) = B + \alpha G^{-1}(Z)$, which was the approach taken in Ref.~\onlinecite{Czarnik_8}. This is also closely related to proposal of Ref.~\onlinecite{Dziarmaga_gradient}, where $Z$ can be thought of as the gradient of the overlap, and $G^{-1}$ can be interpreted as a preconditioner, for which natural choices are given by the full tangent space metric used in Ref.~\onlinecite{Dziarmaga_gradient} or the local environment used here. Indeed, both of these choices have recently been established to perform well as preconditioner in general PEPS optimization \cite{Zhang_preconditioner}.

For the present purpose, however, we compute the gradient of the cost function, already evaluated using CTMRG environments, using the framework of Automatic Differentiation. In that case, the gradient can receive additional contributions originating from the $B$ dependence in the CTMRG computation itself. The resulting gradient is then used in a nonlinear gradient-based optimization algorithm, such as the quasi-Newton algorithm of Broyden–Fletcher–Goldfarb–Shanno with Limited memory (L-BFGS). The automatic differentiation can be performed using the open-source software package PEPSKit.jl~\cite{PEPSKit}. As in the global truncation method from previous subsection, this approach also requires an initial guess from which to start the optimization procedure. Since the evolution operator for small values of $\beta$ is close to the identity operator, the previous tensor $A$ can be used as an initial guess for $B$. An alternative is to use the local truncation explained before as an initial guess, which can furthermore be used to reconfigure the virtual spaces of the truncated PEPO, which is important in the presence of (fermionic) symmetries. Furthermore, during the optimization, reusing the environments from the previous iteration will make the algorithm converge much faster. Indeed, the computational cost of this approach is now dominated by computing the CTMRG environment of the three-layer network, which has a computational complexity that scales as $D_1^6 D_2^3 \chi^3$.

\section{Results and discussion}
\label{sec:results_full}
This section presents our results on using the cluster expansion for determining the phase diagram of a model of interacting fermions on a two-dimensional lattice. The Hamiltonian of this model is introduced in Subsection~\ref{sec:SF}. We then compare the different truncation schemes in Subsection~\ref{sec:comparison_truncation_schemes}, before presenting the full simulation results in Subsection~\ref{sec:phase_diagram}. But first, we benchmark the proposed cluster expansion-based approach in a well-known spin model, as this particular workflow for constructing two-dimensional thermal states had not yet been tested in the original cluster expansion papers \cite{ClusterExpansion1,ClusterExpansion2}.

\subsection{Benchmark of the cluster expansion workflow}
\label{sec:Comparison_CE_Trotter_SU}

In this first subsection, we will benchmark the performance of the cluster expansion in constructing thermal states. As this performance should not depend on the fermionic statistics, we address this question in a well-studied spin model exhibiting a finite-temperature phase transition, namely the quantum Ising model, with its Hamiltonian given by
\begin{equation}
H = - J \sum_{\left<i,j\right>} \sigma_i^Z \sigma_j^Z - g \sum_{i} \sigma_i^X \; .
\end{equation}
The case $g = 2.5$ will be investigated, with expected critical temperature $T_c = 1.2736$ from QMC calculations \cite{QMC_Tc_Ising}. 

We compare the $P=3$ cluster expansion PEPO, exhibiting $D_2=5$  and accurate up to second-order, with a PEPO constructed from the ST decomposition, using identical imaginary time step $\Delta \beta$ and contracted using the same truncation strategy. In particular, we use the local truncation strategy, which was already established to perform well in Ref.~\onlinecite{TTNR}. Note that the quantum Ising model is a favorable case for the ST decomposition, as it admits a $D_2=2$ PEPO representation for the second-order symmetric ST decomposition. For more complicated models, the standard ST decomposition will only be accurate to first order, and the symmetric second-order decomposition will consists of several layers of gates, thus giving rise to a PEPO with higher bond dimension.

\begin{figure}
    \centering
    \begin{subfigure}{\linewidth}
        \includegraphics[width=0.9\columnwidth]{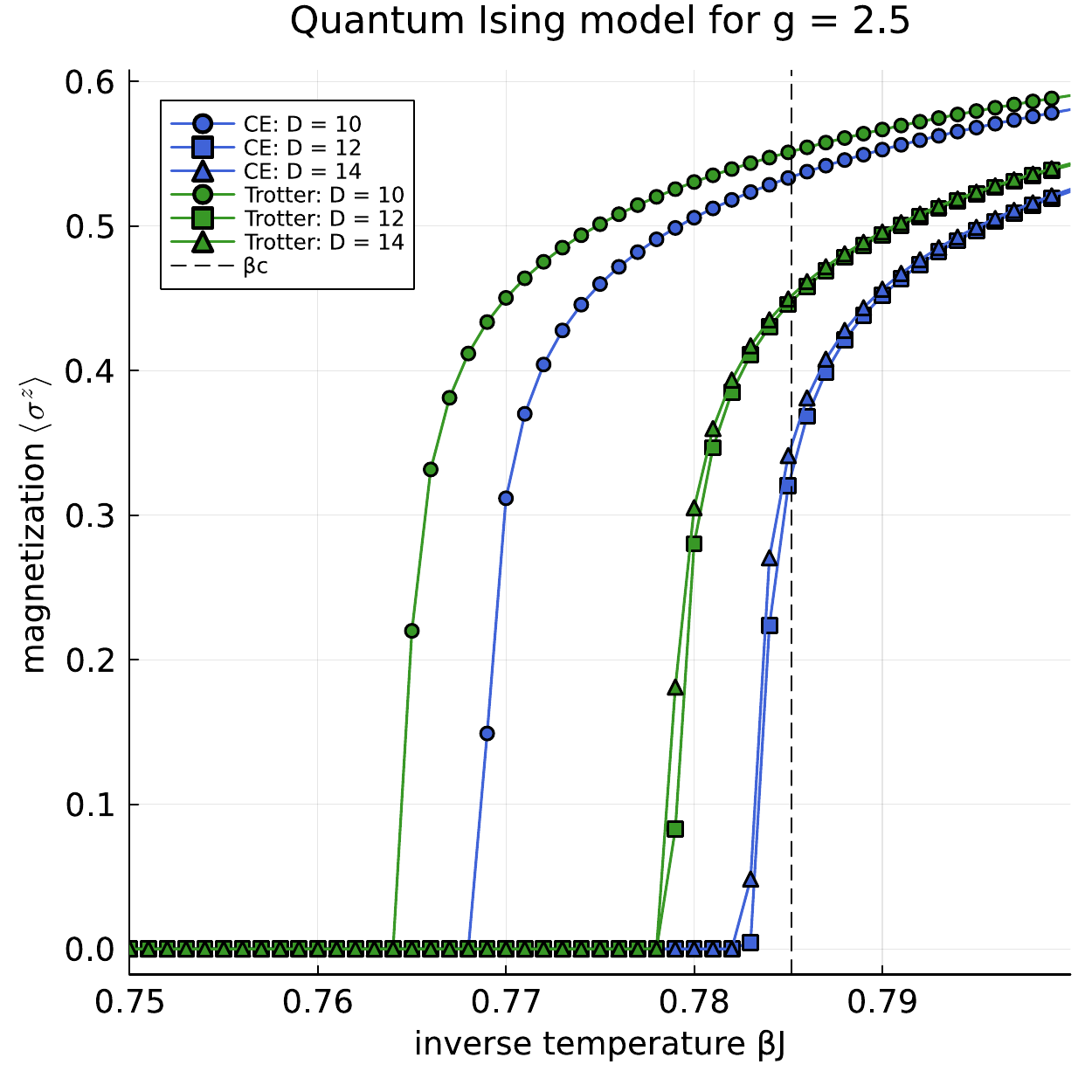}        
        \end{subfigure} 
    \begin{subfigure}{\linewidth}
        \includegraphics[width=0.9\columnwidth]{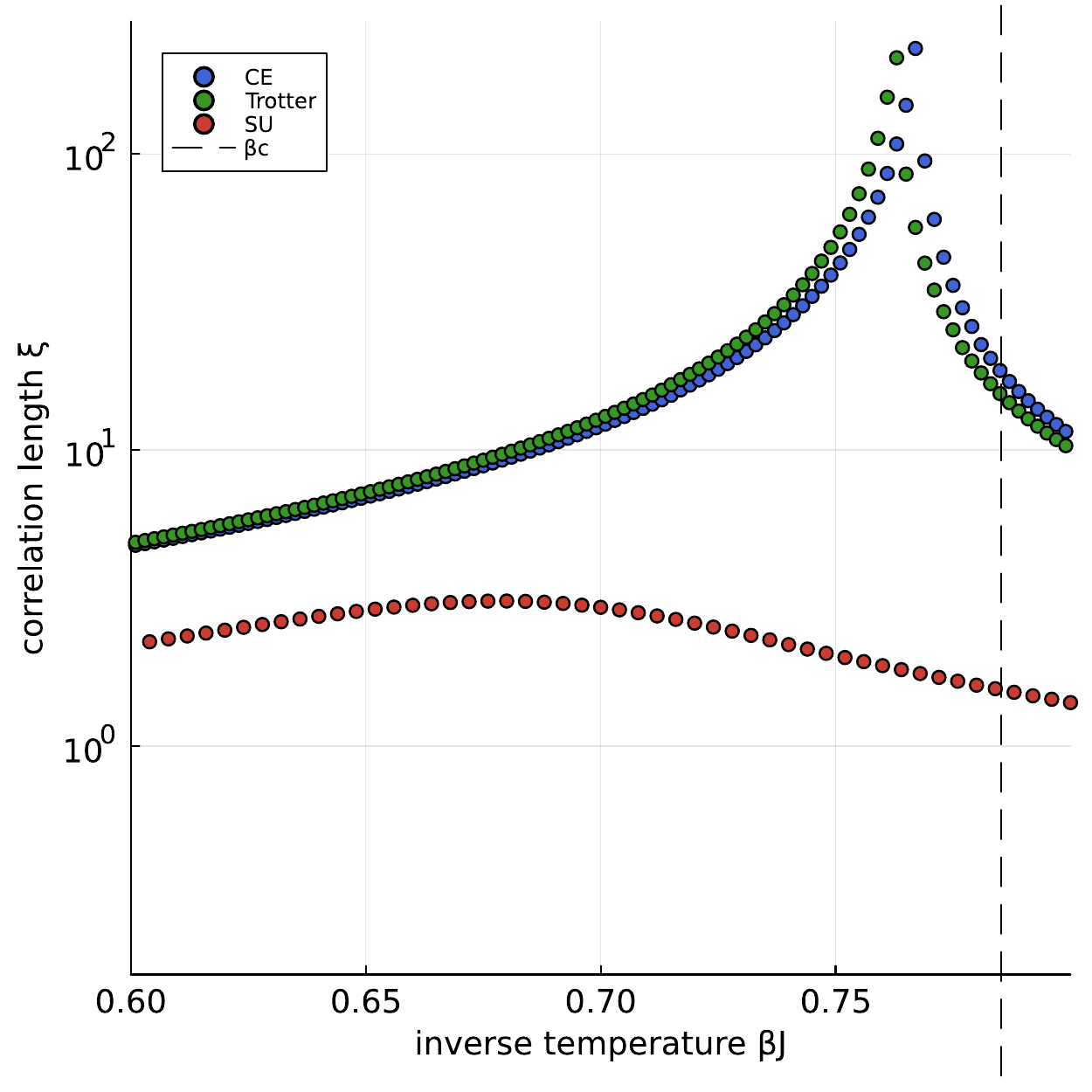}
    \end{subfigure}
    \caption{(top) Magnetization of the quantum Ising model for $g = 2.5$ in function of the inverse temperature for both the second-order ST decomposition and second-order ($P=3$) CE, both using the local truncation scheme. We indicate an accurate estimate $\beta_c = 1 / 1.2736$ of the critical temperature, based on sign-problem-free Monte Carlo results \cite{QMC_Tc_Ising}. (bottom) Correlation length for the same model in comparison with SU for $\Delta \tau = 10^{-5}$ and $D = 4$. The observables were calculated using VUMPS with an environment bond dimension $\chi = 16$. The correlation length is defined in appendix \ref{ap:FCLS}. 
    }
    \label{fig:Quantum_Ising_CE_vs_T}
\end{figure}

Figure \ref{fig:Quantum_Ising_CE_vs_T} presents the magnetization and correlation length, computed from the thermal state obtained from the local truncation method with the ST decomposition and cluster expansion, for different values of the final bond dimension $D_1$. In the bottom panel, depicting the correlation length, we also include results obtained using the SU method. From the magnetization data, we observe that both the cluster expansion and the ST decomposition yield converged results around $D = D_1 = 12$, but that the cluster expansion method gives a significantly more accurate prediction of the critical temperature $T_c$. Even though both are second-order approximations \footnote{While second-order methods exhibit an error that scales as $\mathcal{O}(\Delta \beta^3)$ after a single time step, the accumulated error (ignoring additional truncation errors) to reach a total (imaginary) time $\beta = N \Delta \beta$ scales as $\mathcal{O}(N \Delta \beta^3) = \mathcal{O}(\Delta \beta^2)$.}, the cluster expansion likely includes more higher-order terms which reduce the absolute error. These results were obtained with a time step $\Delta \beta = 10^{-2}$, and we do indeed observe PEPOs resulting from the cluster expansion and the ST decomposition come closer together for smaller time steps. The current value of $\Delta \beta$ strikes a good balance between accuracy and cost. For smaller values of the time step, the number of steps required to reach a given temperature is larger, which results in a higher computational cost and possibly in a higher accumulation of truncation errors.

The correlation length data yields a sharp peak for both the ST decomposition and the cluster expansions, while the peak is smeared out and reaches a much lower maximum for SU, at a temperature that is furthermore significantly shifted with respect to the expected result. Since the peak of the correlation length is often used in a finite entanglement scaling approach to obtain a more accurate estimate of the critical temperature, using SU will yield less accurate results as compared to the other two methods. Furthermore, we observe that the computational cost of performing one step in the imaginary-time evolution with the local truncation method is approximately the same as that of SU. However, SU requires much smaller time steps. Because the calculation of the initial PEPO comes at a negligible cost when using both the ST decomposition and cluster expansions, we conclude that these options are both faster and more accurate. More investigation is needed to compare the scheme with fast full update \cite{FFU}, which is expected to perform better than SU. 

\subsection{\label{sec:SF}Interacting spinless fermions}
We now turn our attention to fermionic system. In particular, we consider spinless fermions hopping on a square lattice with nearest-neighbor interactions. The Hamiltonian describing this system is given by
\begin{eqnarray}
H = &-& t \sum_{\left<i,j\right>} \left(c_i^\dagger c_j + \text{h.c.} \right) \nonumber \\
   &+& V \sum_{\left<i,j\right>} \left(n_i - \frac{1}{2} \right) \left(n_j - \frac{1}{2} \right) - \mu \sum_i n_i,\label{eq:spinlessham}
\end{eqnarray}
where $c_i^\dagger$ ($c_i$) represents the creation (annihilation) operator and $n_i = c_i^\dagger c_i$ the number operator on site $i$. $\left<i,j\right>$ represents nearest neighbors. This Hamiltonian consists of a hopping term, a density-density interaction, which will be attractive for negative values of $V$, and a chemical potential to change the average fermion density.

In this work, we set $\mu = 0$ to have an explicitly particle-hole symmetric Hamiltonian. For $V < 0$, there is an Ising-like transition between a homogeneous phase at high temperatures and the coexistence of high-density and low-density homogeneous phases at low temperatures \cite{SF_mean_field, SF_QMC}. Ref.~\onlinecite{SF_mean_field} also found superconducting order for $V < 0$ and $\beta/t > 25$. This regime is currently still out of reach with the current TN methods and will not be discussed in this paper.

\subsection{Comparison of different truncation schemes}
\label{sec:comparison_truncation_schemes}
\begin{figure}
    \centering
    \begin{subfigure}{\linewidth}
        \includegraphics[width=1\columnwidth]{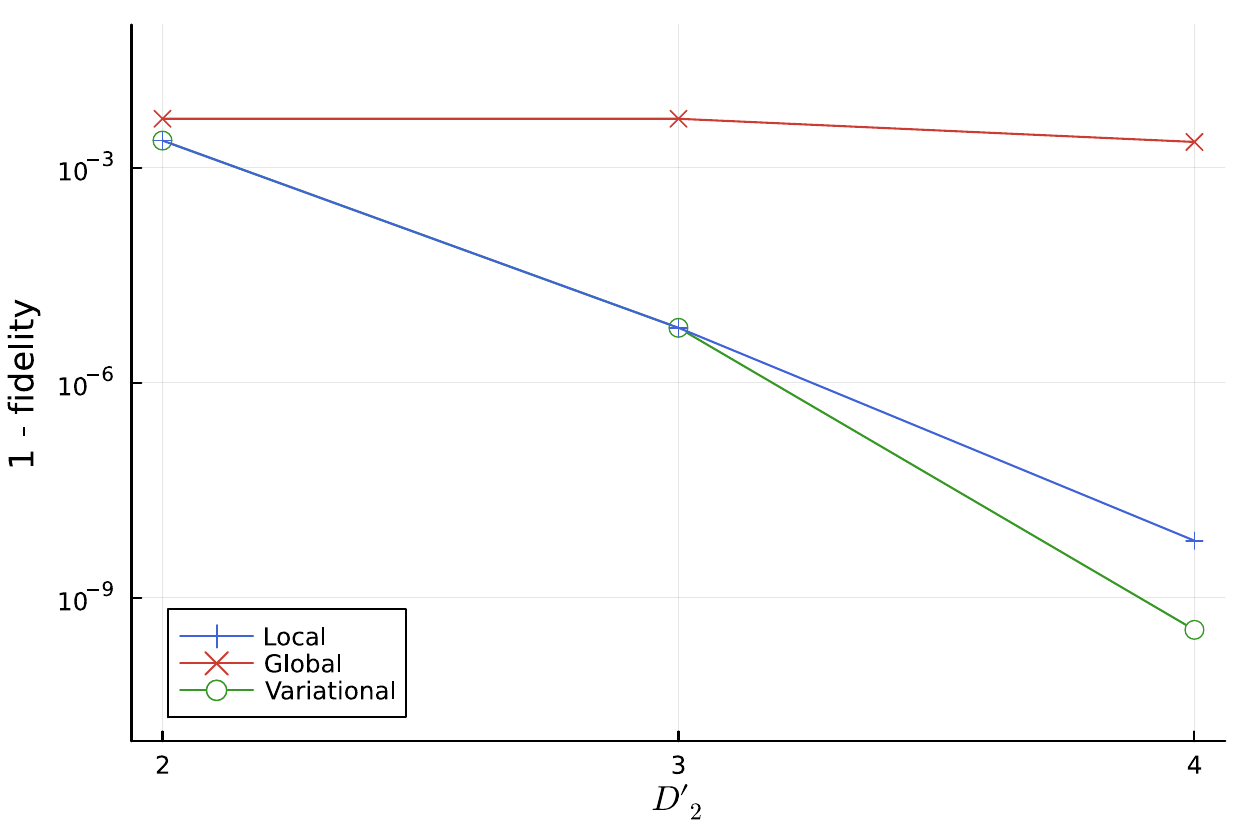}        
        \end{subfigure} 
    \begin{subfigure}{\linewidth}
        \includegraphics[width=1\columnwidth]{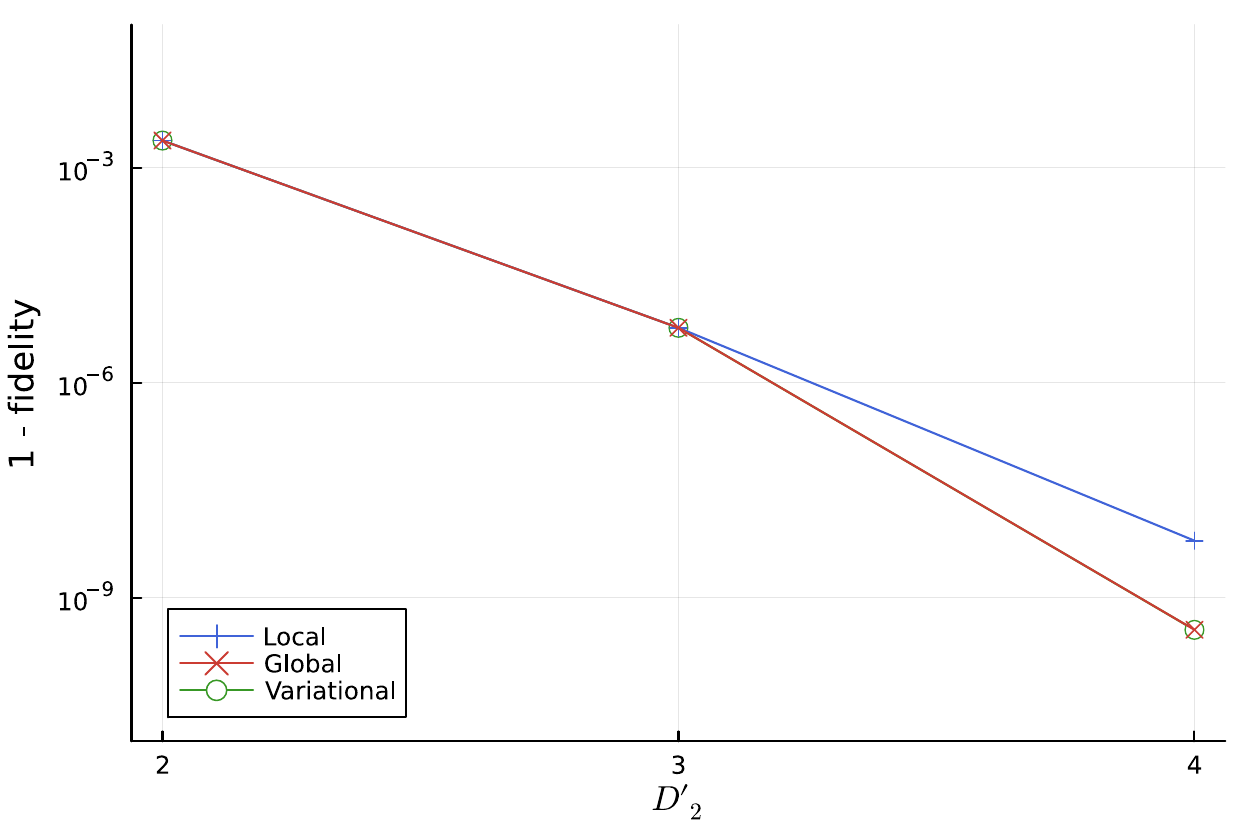}
    \end{subfigure}
    \caption{Fidelity between the exact cluster expansion PEPO of the free spinless fermion model at $\beta = 0.1$ and the PEPO resulting from truncating it from $D_2 = 5$ to a lower bond dimension $D^\prime_2$ using the different truncation schemes of section \ref{sec:truncation_schemes} with a random initial guess (top panel) and the local truncation result as initial guess (bottom panel). Bond dimension $\chi = 20$ was used for the CTMRG environment in both the global and variational truncation scheme. Since the exact PEPO has bond dimensions $5$, truncating to $D^\prime_2 \geq 5$ would result in a fidelity of $1$ for all truncation schemes.}
    \label{fig:Comparison_truncation_schemes}
\end{figure}

Before studying the full phase diagram of the Hamiltonian in Eq.~\eqref{eq:spinlessham}, we use it to compare the different truncation strategies. We construct the $P=3$ cluster expansion PEPO with bond dimension $D_2=5$ for $V = 0.0$ (e.g. the free case) at $\beta = 0.1$. We then truncate this PEPO to a lower bond dimension using the different strategies outlined in Section~\ref{sec:truncation_schemes}. The fidelity between the truncated and the exact PEPO is then calculated as a measure of accuracy. Figure \ref{fig:Comparison_truncation_schemes} shows the results.

All truncation schemes yield higher fidelity for larger bond dimensions, as expected. The fidelity of the variational truncation scheme, for which the fidelity is exactly the cost function, always outperforms the other schemes, irrespective of the chosen initial guess. For $D = 2$ and $D = 3$, however, the difference is very small, which means that in those cases the PEPO resulting from the local truncation scheme is already very close to the optimal one. For $D=4$ on the other hand, the variational truncation can improve the fidelity by an order of magnitude, though in absolute terms also the fidelity of the local truncation is very close to one. The global truncation scheme can find solutions that are very close to the variational truncation, when initialized using the solution from the local truncation. However, this algorithm seems quite sensitive to the choice of initial guess. In particular, global truncation will sometimes converge to a solution which has much lower fidelity than the solution of the local and variational schemes, when the algorithm starts from a poorly chosen initial guess. 

Though the variational approach should thus be used in situations where maximal accuracy is needed, Fig.~\ref{fig:Comparison_truncation_schemes} illustrates that the much cheaper local truncation scheme can already produce a solution with high fidelity, which will yield sufficiently accurate results in most cases. Furthermore, the higher computational cost of the variational approach limits the achievable bond dimension, which can be a problem for low temperatures. For this reason, the local truncation scheme was used in determining the phase diagram of the spinless fermion model in the next subsection of this work. This coincides with the approach that was used in Ref.~\onlinecite{TTNR}. 

\begin{figure}
    \centering
    \includegraphics[width=1\columnwidth]{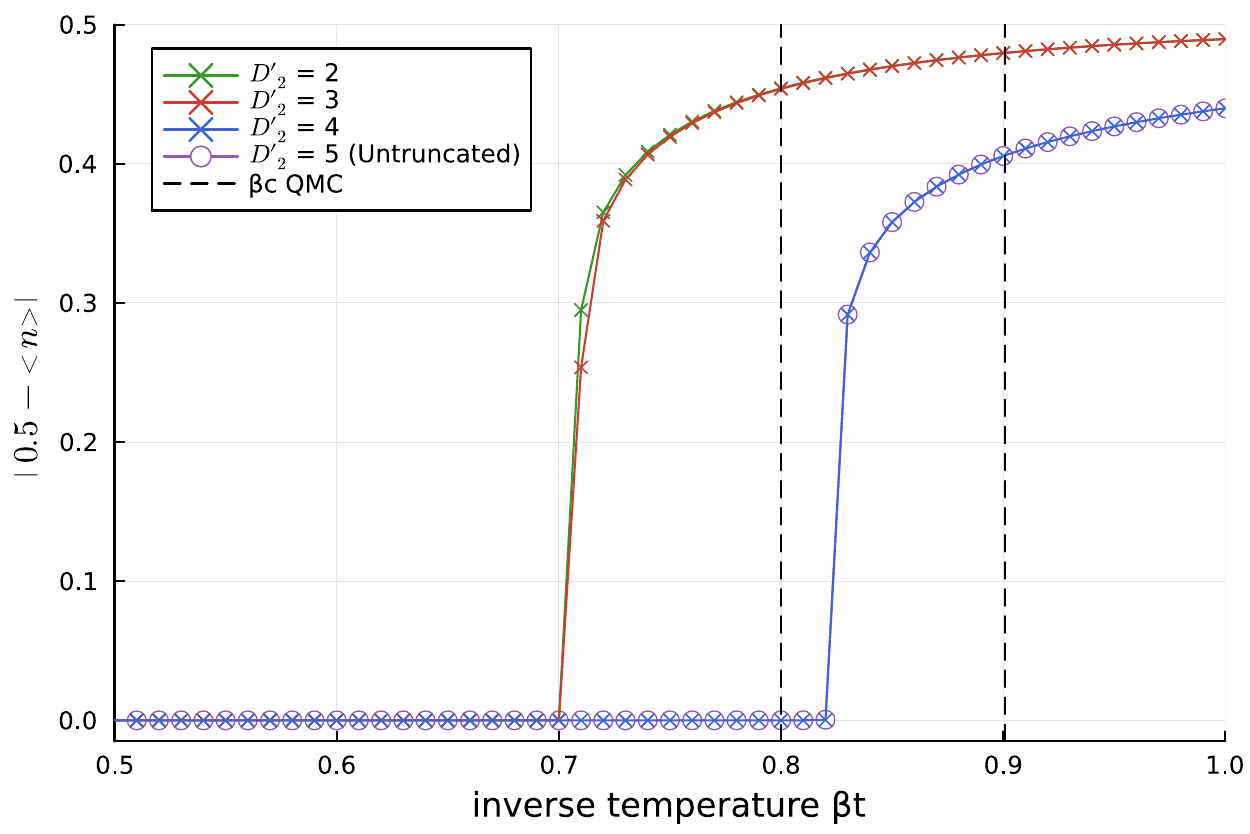}
    \caption{Occupancy of the spinless fermion model for $V = -2.5$ where the update-PEPOs were truncated to different bond dimensions using the local truncation scheme. The subsequent time evolution was performed with the local truncation scheme and $D_1 = 5$. The dotted lines denote the predictions from QMC up to 1 standard deviation \cite{SF_QMC}.}
    \label{fig:Truncated_PEPO_Local}
\end{figure}

The truncation of a single PEPO layer was here only used to illustrate the performance of the different truncation strategies. However, this can be a useful preprocessing step in a full simulation, in order to reduce the computational cost of the calculation, by already compressing the bond dimension $D_2$ of the PEPO representing a single time step $\Delta \tau$. On the square lattice, the bond dimension of the second-order ($P=3$) cluster expansion PEPO of a nearest-neighbor Hamiltonian is equal to $D_2 = 1 + d^2$. While this remains feasible for the spinless fermion model, where $d = 2$ yields $D_2 = 5$, this bond dimension would grow to $D_2 = 10$ for the $t$-$J$ model and $D_2 = 17$ for the Hubbard model, making accurate simulations much more expensive. Figure \ref{fig:Truncated_PEPO_Local} shows the influence of compressing the bond dimension of the initial PEPO for the spinless fermion model at $V = -2.5$ using the local truncation scheme. It can be seen that the estimated critical temperature is not influenced by a truncation from $D = 5$ to $D = 4$, whereas it is significantly shifted by a truncation to lower bond dimensions. The fact that the results for $D = 2$ and $D = 3$ are almost exactly the same is due to the presence of a multiplet, which is broken for $D = 3$. In order to have a fair comparison, the density operator was truncated to the same bond dimension ($D_1 = 5$) after each time step for all calculations. Similar results are obtained if the compression is performed using the variational truncation scheme, which could have been predicted from Fig.~\ref{fig:Comparison_truncation_schemes} by the fact the fidelity between the original and truncated PEPO is very similar for both schemes. Further research is needed to investigate whether the same holds for the $t$-$J$ and the Hubbard model.

\subsection{Finite-temperature phase diagram}
\label{sec:phase_diagram}
We now probe the finite-temperature phase diagram of the spinless fermion model in Eq.~\eqref{eq:spinlessham} at the charge-neutral point for negative (attractive) values of the interaction strength. For a range of different interaction strengths, we construct the finite temperature state for different values $\beta_n = n \Delta\beta$ by using the second-order ($P=3$) cluster expansion PEPO and combining the layers using the local truncation scheme. To determine the exact critical temperature for each interaction strength, we calculate the charge occupancy $n$ in the thermal state, as well as the correlation length $\xi$ of the boundary MPS \cite{Cirac_2008, BoundaryMPS1, VUMPS_2D}. At the phase transition between the homogeneous and the phase-separated region, the charge occupancy will change from the symmetric case ($n = \frac{1}{2}$) to the symmetry-broken case ($n < \frac{1}{2}$ or $n > \frac{1}{2}$), and the correlation length diverges. 

\begin{figure}
    \centering
    \begin{subfigure}{\linewidth}
        \includegraphics[width=1\columnwidth]{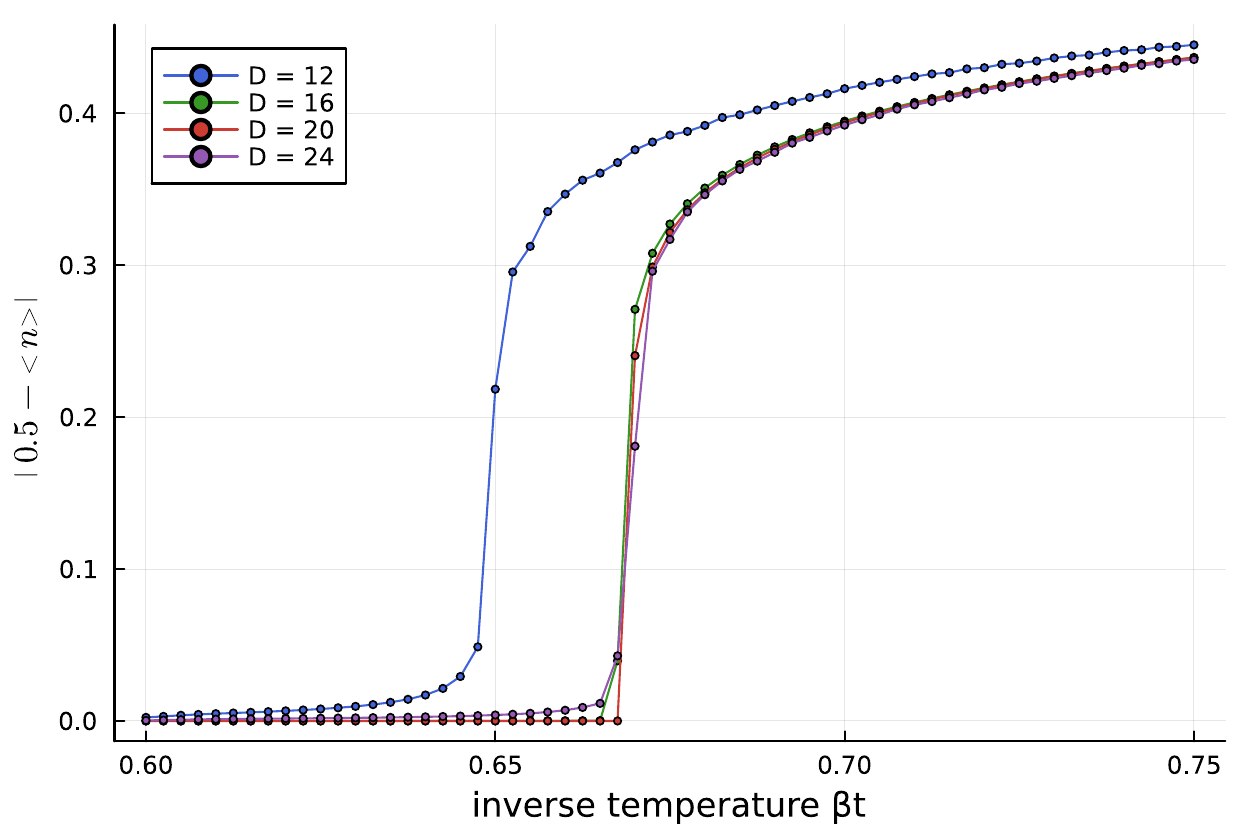}        
        \end{subfigure} 
    \begin{subfigure}{\linewidth}
        \includegraphics[width=1\columnwidth]{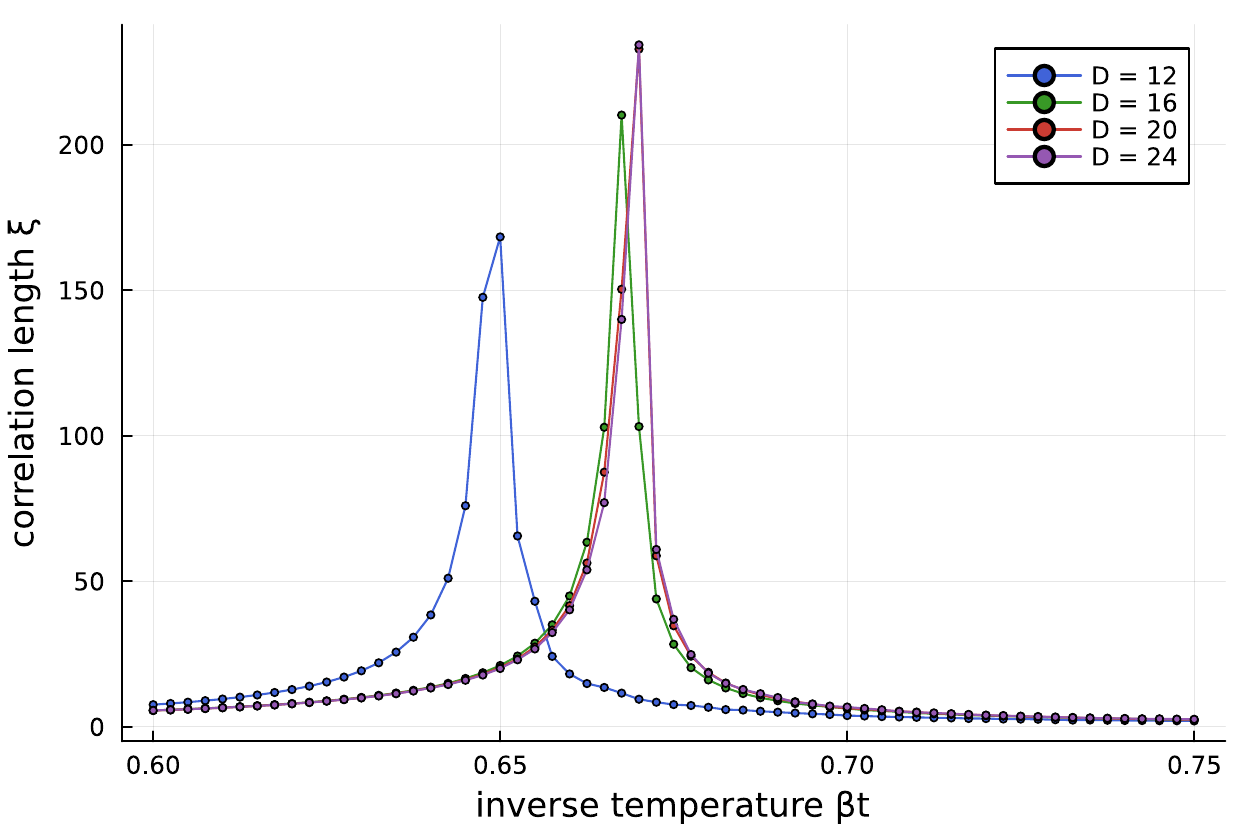}
    \end{subfigure}
    \caption{Occupancy (top) and correlation length (bottom) of the spinless fermion model for $V = -3.0$ for different bond dimensions. The VUMPS algorithm used in the calculation of the observables requires only a single layer of the PEPO.}
    \label{fig:corrlength}
\end{figure}

Fig.~\ref{fig:corrlength} plots the results of these calculations for $V=-3$. Hereto, we have computed the boundary MPS of the final (single-layer) PEPO using the VUMPS algorithm \cite{VUMPS, VUMPS_2D} for different values of $\chi$, and extrapolated the resulting magnetization and correlation length data for $\chi \to \infty$ using finite entanglement scaling \cite{Rams_2018}, which is explained in appendix \ref{ap:FCLS}. Hence, Fig.~\ref{fig:corrlength} shows the extrapolated values $\lim_{\chi \to \infty} \xi$ and $\lim_{\chi \to \infty} \left< n \right>$ for each value of $D$ and $\beta$. The value of $\beta$ that maximizes the correlation length serves as an estimate of the critical temperature $T_c$. Because of the finite PEPO bond dimension, we are essentially simulating an effective Hamiltonian $H_{\text{eff}}$ with fewer quantum fluctuations than the true Hamiltonian $H$. Each of these effective Hamiltonians has a critical inverse temperature $\beta_c$, which thus depends on the bond dimension. Extrapolating $\beta_c(1 / D)$ to infinite PEPO bond dimension allows us to determine the critical temperature of the true Hamiltonian $H$. In Fig.~\ref{fig:corrlength}, it can be observed that $\beta_c$ has already converged for the highest bond dimensions.

\begin{figure}
    \centering
    \includegraphics[width=1\columnwidth]{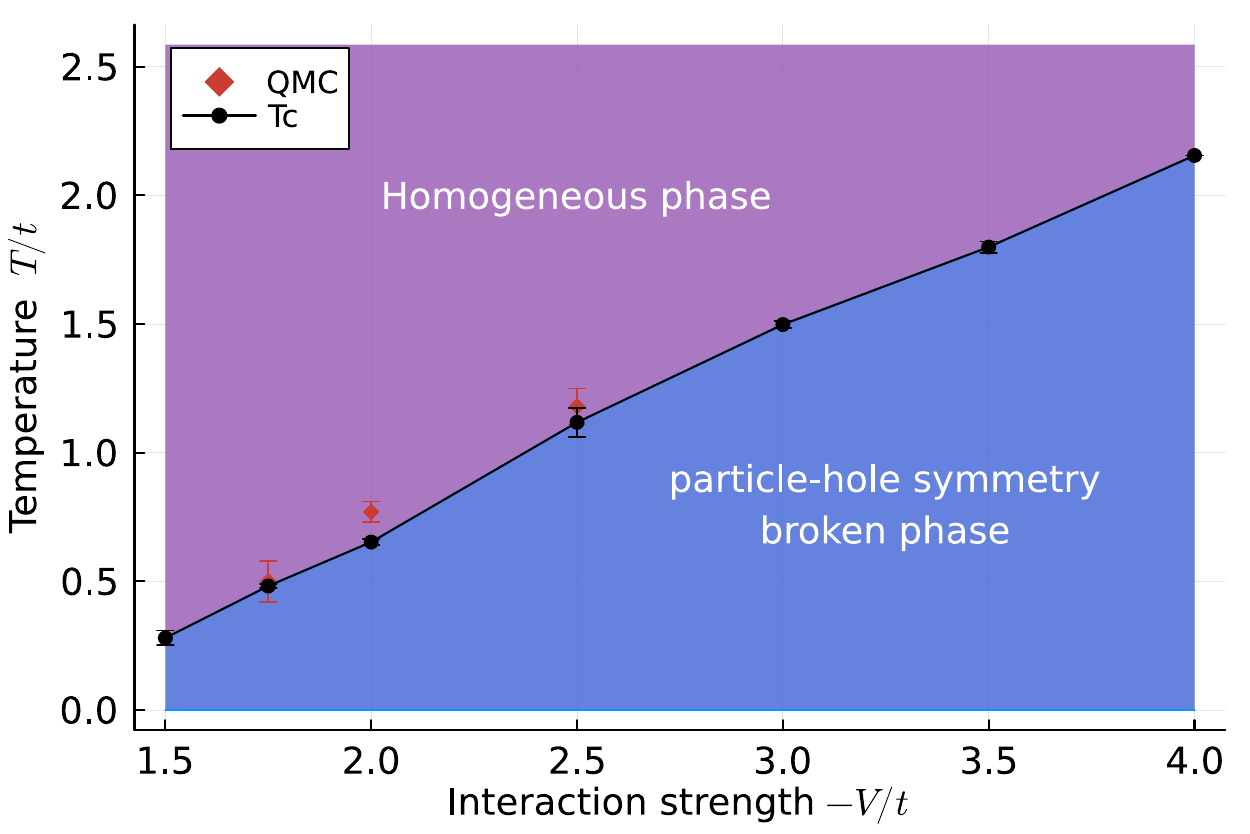}
    \caption{Finite-temperature phase diagram of the spinless fermion model for different values of the interaction strength $V$. $T_c$ denotes the critical temperature as derived from the peak of the correlation function of the extrapolated CE data. The QMC results are taken from Ref.~\onlinecite{SF_QMC}.}
    \label{fig:SF_phase_diagram}
\end{figure}

Fig.~\ref{fig:SF_phase_diagram} depicts the resulting phase diagram of the spinless fermion model with an attractive interaction. The results are in line with the results from QMC in the region where the sign problem is not prohibitively large, aside from a small deviation at the point $V = -2.0$. As the sign problem in QMC intensifies for stronger interactions, no reference data is available in this part of the phase diagram.

\section{Conclusion}\label{sec:conclusion}
In this paper, we have studied the thermal phase transitions of a two-dimensional spinless fermion model with attractive interactions on the square lattice. We showed that using the cluster expansion instead of the ST decomposition to construct PEPO representations of $e^{-\Delta \tau H}$ can yield very accurate results, in particular for determining the critical temperature. For truncating the imaginary-time evolution of the PEPO, we compared local, global, and variational schemes. We found that, even though the variational methods can obtain more accurate results than the local truncation scheme for the same bond dimension, the difference is small and does not justify the significantly higher computational cost. The lower computational and memory cost of the local truncation scheme allows us to use a higher bond dimensions, which is expected to have a net positive effect on the final accuracy of the total workflow. 

The proposed method paves the way for a more accurate simulation of low-temperature physics of interacting two-dimensional fermions, and could be applied to a range of different models, among which are the $t$-$J$ model and the Hubbard model \cite{Dziarmaga_NTU, Sinha2025, ttJ, Variational_PEPS_Finite_T2, PEPS_Finite_T_Hubbard1}. Since these models exhibit a rich phase diagram that necessitates very accurate truncation schemes, further research is needed to investigate whether the proposed methods are sufficient to capture the relevant phase transitions. Hereto, it could be beneficial to investigate additional truncation schemes that are are intermediate between the local and variational schemes in terms of both accuracy and cost. Indeed, generalizing the different variations of the Neighborhood Tensor Update \cite{Zhang, Dziarmaga_NTU, Sinha2025} to the context of cluster expansions might be worthwile. In order to improve the accuracy of the cluster expansion PEPO itself, an alternative to increasing the maximal cluster size $P$, is to compose several second-order ($P=3$) PEPOs with well chosen fractions of the step size $\Delta \tau$ to obtain a fourth-order PEPO \cite{Higher_order_hybrid}, the bond dimension of which can potentially be further compressed using one of the truncation schemes. This would result in a higher-order method while circumventing the exponential increase of the bond dimension of the cluster expansion.

Furthermore, the construction of the cluster expansion is not restricted to the square lattice. Given the diversity of interesting models and phases that exist on different lattices, such as the triangular, honeycomb, and Kagome lattice, it would be natural to extend the cluster expansion approach to these lattices. Because of the additional geometric frustration that can be present, extra care will be needed to make sure that the environment calculation of the partition function is well defined and converges \cite{Frustration_Vanhecke_Colbois}. Another application of the proposed methodology could be real-time evolution, which amounts to the substitution $\Delta \beta \to i \Delta t$ in the expansion. Even though higher bond dimensions will probably be needed in the truncation schemes, since high-energy contributions are not exponentially suppressed, the cluster expansion method could again prove to be a more accurate alternative to using the ST decomposition.

\textit{Code availability.} The code that was used to obtain the results in this paper is available in our open source \textit{ClusterExpansions} library \cite{Code_ClusterExpansions}. This includes the construction of the cluster expansion PEPO for any order up to $P = 9$ for different models (Ising, Heisenberg, spinless fermions, $t$-$J$, and Hubbard) on both the square and triangular lattice, together with the implementation of the different truncation schemes. Questions about or suggestions for the further development of this code are much appreciated.

\begin{acknowledgments}
We thank Bram Vanhecke, David Devoogdt, Frank Verstraete, Jacek Dziarmaga, and all members of the Quantum Group Ghent for interesting discussions. The computational resources (Stevin Supercomputer Infrastructure) and services used in this work were provided by the Flemish Supercomputer Center (VSC), funded by Ghent University, the Research Foundation Flanders (FWO), and the Flemish Government. S.D.M. is supported by the FWO under doctoral fellowship No. 11A3C25N. A.U. is supported by FWO Junior Postdoctoral Fellowship (grant No. 3E0.2025.0049.01) and Watanabe Foundation. This research is supported by the European Research Council (ERC) under the European Union's Horizon 2020 program (grant agreement No. 101125822-GaMaTeN and No. 101076597-SIESS).
\end{acknowledgments}

\bibliography{Ref}

@article{ClusterExpansion1,
  title = {Symmetric cluster expansions with tensor networks},
  author = {Vanhecke, Bram and Vanderstraeten, Laurens and Verstraete, Frank},
  journal = {Phys. Rev. A},
  volume = {103},
  issue = {2},
  pages = {L020402},
  numpages = {5},
  year = {2021},
  month = {Feb},
  publisher = {American Physical Society},
  doi = {10.1103/PhysRevA.103.L020402},
  url = {https://link.aps.org/doi/10.1103/PhysRevA.103.L020402}
}

@article{ClusterExpansion2,
	title={{Simulating thermal density operators with cluster expansions and tensor networks}},
	author={Bram Vanhecke and David Devoogdt and Frank Verstraete and Laurens Vanderstraeten},
	journal={SciPost Phys.},
	volume={14},
	pages={085},
	year={2023},
	publisher={SciPost},
	doi={10.21468/SciPostPhys.14.4.085},
	url={https://scipost.org/10.21468/SciPostPhys.14.4.085},
}

@article{BTRG,
  title = {Boundary tensor renormalization group},
  author = {Iino, Shumpei and Morita, Satoshi and Kawashima, Naoki},
  journal = {Phys. Rev. B},
  volume = {100},
  issue = {3},
  pages = {035449},
  numpages = {8},
  year = {2019},
  month = {Jul},
  publisher = {American Physical Society},
  doi = {10.1103/PhysRevB.100.035449},
  url = {https://link.aps.org/doi/10.1103/PhysRevB.100.035449}
}

@article{Czarnik_1,
  title = {Projected entangled pair states at finite temperature: Imaginary time evolution with ancillas},
  author = {Czarnik, Piotr and Cincio, Lukasz and Dziarmaga, Jacek},
  journal = {Phys. Rev. B},
  volume = {86},
  issue = {24},
  pages = {245101},
  numpages = {6},
  year = {2012},
  month = {Dec},
  publisher = {American Physical Society},
  doi = {10.1103/PhysRevB.86.245101},
  url = {https://link.aps.org/doi/10.1103/PhysRevB.86.245101}
}

@article{Czarnik_2,
  title = {Fermionic projected entangled pair states at finite temperature},
  author = {Czarnik, Piotr and Dziarmaga, Jacek},
  journal = {Phys. Rev. B},
  volume = {90},
  issue = {3},
  pages = {035144},
  numpages = {7},
  year = {2014},
  month = {Jul},
  publisher = {American Physical Society},
  doi = {10.1103/PhysRevB.90.035144},
  url = {https://link.aps.org/doi/10.1103/PhysRevB.90.035144}
}

@article{Czarnik_3,
  title = {Projected entangled pair states at finite temperature: Iterative self-consistent bond renormalization for exact imaginary time evolution},
  author = {Czarnik, Piotr and Dziarmaga, Jacek},
  journal = {Phys. Rev. B},
  volume = {92},
  issue = {3},
  pages = {035120},
  numpages = {8},
  year = {2015},
  month = {Jul},
  publisher = {American Physical Society},
  doi = {10.1103/PhysRevB.92.035120},
  url = {https://link.aps.org/doi/10.1103/PhysRevB.92.035120}
}

@article{Czarnik_4,
  title = {Variational approach to projected entangled pair states at finite temperature},
  author = {Czarnik, Piotr and Dziarmaga, Jacek},
  journal = {Phys. Rev. B},
  volume = {92},
  issue = {3},
  pages = {035152},
  numpages = {8},
  year = {2015},
  month = {Jul},
  publisher = {American Physical Society},
  doi = {10.1103/PhysRevB.92.035152},
  url = {https://link.aps.org/doi/10.1103/PhysRevB.92.035152}
}

@article{SF_mean_field,
  title = {Interacting spinless fermions on the square lattice: Charge order, phase separation, and superconductivity},
  author = {Ma, Kou-Han and Tong, Ning-Hua},
  journal = {Phys. Rev. B},
  volume = {104},
  issue = {15},
  pages = {155116},
  numpages = {14},
  year = {2021},
  month = {Oct},
  publisher = {American Physical Society},
  doi = {10.1103/PhysRevB.104.155116},
  url = {https://link.aps.org/doi/10.1103/PhysRevB.104.155116}
}

@article{Rams_2018,
  title = {Precise Extrapolation of the Correlation Function Asymptotics in Uniform Tensor Network States with Application to the Bose-Hubbard and XXZ Models},
  author = {Rams, Marek M. and Czarnik, Piotr and Cincio, Lukasz},
  journal = {Phys. Rev. X},
  volume = {8},
  issue = {4},
  pages = {041033},
  numpages = {19},
  year = {2018},
  month = {Nov},
  publisher = {American Physical Society},
  doi = {10.1103/PhysRevX.8.041033},
  url = {https://link.aps.org/doi/10.1103/PhysRevX.8.041033}
}

@misc{QMC_sign_problem,
      title={Asymptotic sign free in interacting fermion models}, 
      author={Zi-Xiang Li and Zhou-Quan Wan and Hong Yao},
      year={2022},
      eprint={2211.00663},
      archivePrefix={arXiv},
      primaryClass={cond-mat.str-el},
      url={https://arxiv.org/abs/2211.00663}, 
}

@article{SF_QMC,
  title = {Two-dimensional spin-polarized fermion lattice gases},
  author = {Gubernatis, J. E. and Scalapino, D. J. and Sugar, R. L. and Toussaint, W. D.},
  journal = {Phys. Rev. B},
  volume = {32},
  issue = {1},
  pages = {103--116},
  numpages = {0},
  year = {1985},
  month = {Jul},
  publisher = {American Physical Society},
  doi = {10.1103/PhysRevB.32.103},
  url = {https://link.aps.org/doi/10.1103/PhysRevB.32.103}
}

@article{SF_SC_Girvin_1978,
doi = {10.1088/0022-3719/11/18/002},
url = {https://dx.doi.org/10.1088/0022-3719/11/18/002},
year = {1978},
month = {sep},
publisher = {},
volume = {11},
number = {18},
pages = {L751},
author = {S M Girvin},
title = {Critical conductivity of the lattice gas},
journal = {Journal of Physics C: Solid State Physics}
}

@article{SU_Jahromi_2019,
   title={Universal tensor-network algorithm for any infinite lattice},
   volume={99},
   ISSN={2469-9969},
   url={http://dx.doi.org/10.1103/PhysRevB.99.195105},
   DOI={10.1103/physrevb.99.195105},
   number={19},
   journal={Physical Review B},
   publisher={American Physical Society (APS)},
   author={Jahromi, Saeed S. and Orús, Román},
   year={2019},
   month=may }

@article{Maarten_Taylor,
	title={{Efficient higher-order matrix product operators for time evolution}},
	author={Maarten Van Damme and Jutho Haegeman and Ian McCulloch and Laurens Vanderstraeten},
	journal={SciPost Phys.},
	volume={17},
	pages={135},
	year={2024},
	publisher={SciPost},
	doi={10.21468/SciPostPhys.17.5.135},
	url={https://scipost.org/10.21468/SciPostPhys.17.5.135},
}

@article{FCLS_Corboz_Czarnik,
  title = {Finite Correlation Length Scaling with Infinite Projected Entangled-Pair States},
  author = {Corboz, Philippe and Czarnik, Piotr and Kapteijns, Geert and Tagliacozzo, Luca},
  journal = {Phys. Rev. X},
  volume = {8},
  issue = {3},
  pages = {031031},
  numpages = {10},
  year = {2018},
  month = {Jul},
  publisher = {American Physical Society},
  doi = {10.1103/PhysRevX.8.031031},
  url = {https://link.aps.org/doi/10.1103/PhysRevX.8.031031}
}

@article{Variational_PEPS_Finite_T2,
  title = {Variational tensor network renormalization in imaginary time: Benchmark results in the Hubbard model at finite temperature},
  author = {Czarnik, Piotr and Rams, Marek M. and Dziarmaga, Jacek},
  journal = {Phys. Rev. B},
  volume = {94},
  issue = {23},
  pages = {235142},
  numpages = {12},
  year = {2016},
  month = {Dec},
  publisher = {American Physical Society},
  doi = {10.1103/PhysRevB.94.235142},
  url = {https://link.aps.org/doi/10.1103/PhysRevB.94.235142}
}

@article{TN1,
author = {F. Verstraete and V. Murg and J.I. Cirac},
title = {Matrix product states, projected entangled pair states, and variational renormalization group methods for quantum spin systems},
journal = {Advances in Physics},
volume = {57},
number = {2},
pages = {143--224},
year = {2008},
publisher = {Taylor \& Francis},
doi = {10.1080/14789940801912366},
URL = {https://doi.org/10.1080/14789940801912366},
eprint = {https://doi.org/10.1080/14789940801912366
}
}

@article{TN2,
title = {The density-matrix renormalization group in the age of matrix product states},
journal = {Annals of Physics},
volume = {326},
number = {1},
pages = {96-192},
year = {2011},
note = {January 2011 Special Issue},
issn = {0003-4916},
doi = {https://doi.org/10.1016/j.aop.2010.09.012},
url = {https://www.sciencedirect.com/science/article/pii/S0003491610001752},
author = {Ulrich Schollwöck}
}

@article{TN3,
title = {A practical introduction to tensor networks: Matrix product states and projected entangled pair states},
journal = {Annals of Physics},
volume = {349},
pages = {117-158},
year = {2014},
issn = {0003-4916},
doi = {https://doi.org/10.1016/j.aop.2014.06.013},
url = {https://www.sciencedirect.com/science/article/pii/S0003491614001596},
author = {Román Orús}
}

@article{PEPS_Finite_T_Hubbard1,
  title = {Finite-temperature tensor network study of the Hubbard model on an infinite square lattice},
  author = {Sinha, Aritra and Rams, Marek M. and Czarnik, Piotr and Dziarmaga, Jacek},
  journal = {Phys. Rev. B},
  volume = {106},
  issue = {19},
  pages = {195105},
  numpages = {16},
  year = {2022},
  month = {Nov},
  publisher = {American Physical Society},
  doi = {10.1103/PhysRevB.106.195105},
  url = {https://link.aps.org/doi/10.1103/PhysRevB.106.195105}
}

@article{SF_SC1,
  title = {Correlation exponents and the metal-insulator transition in the one-dimensional Hubbard model},
  author = {Schulz, H. J.},
  journal = {Phys. Rev. Lett.},
  volume = {64},
  issue = {23},
  pages = {2831--2834},
  numpages = {0},
  year = {1990},
  month = {Jun},
  publisher = {American Physical Society},
  doi = {10.1103/PhysRevLett.64.2831},
  url = {https://link.aps.org/doi/10.1103/PhysRevLett.64.2831}
}

@article{SF_SC2,
  title = {Exact solution of the strong coupling $t\ensuremath{-}V$ model with twisted boundary conditions},
  author = {Dias, R. G.},
  journal = {Phys. Rev. B},
  volume = {62},
  issue = {12},
  pages = {7791--7801},
  numpages = {0},
  year = {2000},
  month = {Sep},
  publisher = {American Physical Society},
  doi = {10.1103/PhysRevB.62.7791},
  url = {https://link.aps.org/doi/10.1103/PhysRevB.62.7791}
}

@article{SF_SC3,
  title = {Quasi-one-dimensional dynamics and nematic phases in the two-dimensional Emery model},
  author = {Kivelson, Steven A. and Fradkin, Eduardo and Geballe, Theodore H.},
  journal = {Phys. Rev. B},
  volume = {69},
  issue = {14},
  pages = {144505},
  numpages = {7},
  year = {2004},
  month = {Apr},
  publisher = {American Physical Society},
  doi = {10.1103/PhysRevB.69.144505},
  url = {https://link.aps.org/doi/10.1103/PhysRevB.69.144505}
}

@article{SF_SC4_1,
  title = {Stripes and holes in a two-dimensional model of spinless fermions or hardcore bosons},
  author = {Zhang, N. G. and Henley, C. L.},
  journal = {Phys. Rev. B},
  volume = {68},
  issue = {1},
  pages = {014506},
  numpages = {22},
  year = {2003},
  month = {Jul},
  publisher = {American Physical Society},
  doi = {10.1103/PhysRevB.68.014506},
  url = {https://link.aps.org/doi/10.1103/PhysRevB.68.014506}
}

@article{SF_SC4_2,
  title = {Dilute limit of a strongly-interacting model of spinless fermions and hardcore bosons on the square lattice},
  author = {Zhang, N. G. and Henley, C. L.},
  journal = {The European Physical Journal B - Condensed Matter and Complex Systems},
  volume = {38},
  issue = {3},
  year = {2004},
  month = {Apr},
  publisher = {American Physical Society},
  doi = {10.1140/epjb/e2004-00135-8},
}

@misc{PEPS_original,
      title={Renormalization algorithms for Quantum-Many Body Systems in two and higher dimensions}, 
      author={F. Verstraete and J. I. Cirac},
      year={2004},
      eprint={cond-mat/0407066},
      archivePrefix={arXiv},
      primaryClass={cond-mat.str-el},
      url={https://arxiv.org/abs/cond-mat/0407066}, 
}

@article{Finite_T_before_Czarnik1,
  title = {Coarse-graining renormalization by higher-order singular value decomposition},
  author = {Xie, Z. Y. and Chen, J. and Qin, M. P. and Zhu, J. W. and Yang, L. P. and Xiang, T.},
  journal = {Phys. Rev. B},
  volume = {86},
  issue = {4},
  pages = {045139},
  numpages = {9},
  year = {2012},
  month = {Jul},
  publisher = {American Physical Society},
  doi = {10.1103/PhysRevB.86.045139},
  url = {https://link.aps.org/doi/10.1103/PhysRevB.86.045139}
}

@article{Finite_T_before_Czarnik2,
  title = {Exploring corner transfer matrices and corner tensors for the classical simulation of quantum lattice systems},
  author = {Or\'us, Rom\'an},
  journal = {Phys. Rev. B},
  volume = {85},
  issue = {20},
  pages = {205117},
  numpages = {22},
  year = {2012},
  month = {May},
  publisher = {American Physical Society},
  doi = {10.1103/PhysRevB.85.205117},
  url = {https://link.aps.org/doi/10.1103/PhysRevB.85.205117}
}

@article{QMC_Tc_Ising,
  title = {Thermal Ising transitions in the vicinity of two-dimensional quantum critical points},
  author = {Hesselmann, S. and Wessel, S.},
  journal = {Phys. Rev. B},
  volume = {93},
  issue = {15},
  pages = {155157},
  numpages = {11},
  year = {2016},
  month = {Apr},
  publisher = {American Physical Society},
  doi = {10.1103/PhysRevB.93.155157},
  url = {https://link.aps.org/doi/10.1103/PhysRevB.93.155157}
}

@misc{TTNR,
      title={Global Tensor Network Renormalization for 2D Quantum systems: A new window to probe universal data from thermal transitions}, 
        author = {Atsushi Ueda and {S. De Meyer} and Adwait Naravane and Victor Vanthilt and Frank Verstraete},
      year={2025},
      eprint={2508.05406},
      archivePrefix={arXiv},
      primaryClass={cond-mat.str-el},
      url={https://arxiv.org/abs/2508.05406}, 
}

@article{SU_original,
  title = {Accurate Determination of Tensor Network State of Quantum Lattice Models in Two Dimensions},
  author = {Jiang, H. C. and Weng, Z. Y. and Xiang, T.},
  journal = {Phys. Rev. Lett.},
  volume = {101},
  issue = {9},
  pages = {090603},
  numpages = {4},
  year = {2008},
  month = {Aug},
  publisher = {American Physical Society},
  doi = {10.1103/PhysRevLett.101.090603},
  url = {https://link.aps.org/doi/10.1103/PhysRevLett.101.090603}
}

@article{SU_Banuls,
  title = {Algorithms for finite projected entangled pair states},
  author = {Lubasch, Michael and Cirac, J. Ignacio and Ba\~nuls, Mari-Carmen},
  journal = {Phys. Rev. B},
  volume = {90},
  issue = {6},
  pages = {064425},
  numpages = {16},
  year = {2014},
  month = {Aug},
  publisher = {American Physical Society},
  doi = {10.1103/PhysRevB.90.064425},
  url = {https://link.aps.org/doi/10.1103/PhysRevB.90.064425}
}

@article{FFU,
  title = {Infinite projected entangled pair states algorithm improved: Fast full update and gauge fixing},
  author = {Phien, Ho N. and Bengua, Johann A. and Tuan, Hoang D. and Corboz, Philippe and Or\'us, Rom\'an},
  journal = {Phys. Rev. B},
  volume = {92},
  issue = {3},
  pages = {035142},
  numpages = {12},
  year = {2015},
  month = {Jul},
  publisher = {American Physical Society},
  doi = {10.1103/PhysRevB.92.035142},
  url = {https://link.aps.org/doi/10.1103/PhysRevB.92.035142}
}

@article{CTMRG1,
author = {Nishino ,Tomotoshi and Okunishi ,Kouichi},
title = {Corner Transfer Matrix Renormalization Group Method},
journal = {Journal of the Physical Society of Japan},
volume = {65},
number = {4},
pages = {891-894},
year = {1996},
doi = {10.1143/JPSJ.65.891},
URL = {https://doi.org/10.1143/JPSJ.65.891},
eprint = {https://doi.org/10.1143/JPSJ.65.891}
}

@article{CTMRG2,
author = {Nishino ,Tomotoshi and Okunishi ,Kouichi},
title = {Corner Transfer Matrix Algorithm for Classical    Renormalization Group},
journal = {Journal of the Physical Society of Japan},
volume = {66},
number = {10},
pages = {3040-3047},
year = {1997},
doi = {10.1143/JPSJ.66.3040},
URL = {https://doi.org/10.1143/JPSJ.66.3040},
eprint = {https://doi.org/10.1143/JPSJ.66.3040}
}

@article{CTMRG3,
  title = {Simulation of two-dimensional quantum systems on an infinite lattice revisited: Corner transfer matrix for tensor contraction},
  author = {Or\'us, Rom\'an and Vidal, Guifr\'e},
  journal = {Phys. Rev. B},
  volume = {80},
  issue = {9},
  pages = {094403},
  numpages = {4},
  year = {2009},
  month = {Sep},
  publisher = {American Physical Society},
  doi = {10.1103/PhysRevB.80.094403},
  url = {https://link.aps.org/doi/10.1103/PhysRevB.80.094403}
}

@article{Cirac_2008,
  title = {Classical Simulation of Infinite-Size Quantum Lattice Systems in Two Spatial Dimensions},
  author = {Jordan, J. and Or\'us, R. and Vidal, G. and Verstraete, F. and Cirac, J. I.},
  journal = {Phys. Rev. Lett.},
  volume = {101},
  issue = {25},
  pages = {250602},
  numpages = {4},
  year = {2008},
  month = {Dec},
  publisher = {American Physical Society},
  doi = {10.1103/PhysRevLett.101.250602},
  url = {https://link.aps.org/doi/10.1103/PhysRevLett.101.250602}
}

@article{VUMPS,
  title = {Variational optimization algorithms for uniform matrix product states},
  author = {Zauner-Stauber, V. and Vanderstraeten, L. and Fishman, M. T. and Verstraete, F. and Haegeman, J.},
  journal = {Phys. Rev. B},
  volume = {97},
  issue = {4},
  pages = {045145},
  numpages = {31},
  year = {2018},
  month = {Jan},
  publisher = {American Physical Society},
  doi = {10.1103/PhysRevB.97.045145},
  url = {https://link.aps.org/doi/10.1103/PhysRevB.97.045145}
}

@article{VUMPS_2D,
  title = {Faster methods for contracting infinite two-dimensional tensor networks},
  author = {Fishman, M. T. and Vanderstraeten, L. and Zauner-Stauber, V. and Haegeman, J. and Verstraete, F.},
  journal = {Phys. Rev. B},
  volume = {98},
  issue = {23},
  pages = {235148},
  numpages = {17},
  year = {2018},
  month = {Dec},
  publisher = {American Physical Society},
  doi = {10.1103/PhysRevB.98.235148},
  url = {https://link.aps.org/doi/10.1103/PhysRevB.98.235148}
}

@article{Trotter1959,
  author  = {Trotter, H. F.},
  title   = {On the product of semi-groups of operators},
  journal = {Proceedings of the American Mathematical Society},
  volume  = {10},
  pages   = {545},
  year    = {1959},
  doi     = {10.1090/S0002-9939-1959-0108732-6}
}

@article{Suzuki1976,
  author  = {Suzuki, M.},
  title   = {Generalized Trotter’s formula and systematic approximants of exponential operators and inner derivations with applications to many-body problems},
  journal = {Communications in Mathematical Physics},
  volume  = {51},
  pages   = {183},
  year    = {1976},
  doi     = {10.1007/BF01609348}
}

@article{BoundaryMPS1,
  title = {Gradient methods for variational optimization of projected entangled-pair states},
  author = {Vanderstraeten, Laurens and Haegeman, Jutho and Corboz, Philippe and Verstraete, Frank},
  journal = {Phys. Rev. B},
  volume = {94},
  issue = {15},
  pages = {155123},
  numpages = {14},
  year = {2016},
  month = {Oct},
  publisher = {American Physical Society},
  doi = {10.1103/PhysRevB.94.155123},
  url = {https://link.aps.org/doi/10.1103/PhysRevB.94.155123}
}

@software{Code_ClusterExpansions,
  author  = {De Meyer, Sander},
  title   = {ClusterExpansions},
  year    = {2024},
  url     = {https://github.com/sanderdemeyer/ClusterExpansions},
    note = {https://github.com/sanderdemeyer/ClusterExpansions}
}

@article{Fermion_Nick,
  title = {Fermionic matrix product states and one-dimensional topological phases},
  author = {Bultinck, Nick and Williamson, Dominic J. and Haegeman, Jutho and Verstraete, Frank},
  journal = {Phys. Rev. B},
  volume = {95},
  issue = {7},
  pages = {075108},
  numpages = {20},
  year = {2017},
  month = {Feb},
  publisher = {American Physical Society},
  doi = {10.1103/PhysRevB.95.075108},
  url = {https://link.aps.org/doi/10.1103/PhysRevB.95.075108}
}

@article{Mortier_2025,
   title={Fermionic tensor network methods},
   volume={18},
   ISSN={2542-4653},
   url={http://dx.doi.org/10.21468/SciPostPhys.18.1.012},
   DOI={10.21468/scipostphys.18.1.012},
   number={1},
   journal={SciPost Physics},
   publisher={Stichting SciPost},
   author={Mortier, Quinten and Devos, Lukas and Burgelman, Lander and Vanhecke, Bram and Bultinck, Nick and Verstraete, Frank and Haegeman, Jutho and Vanderstraeten, Laurens},
   year={2025},
   month=jan }

@misc{TensorKit,
      title={TensorKit.jl: A Julia package for large-scale tensor computations, with a hint of category theory}, 
      author={Lukas Devos and Jutho Haegeman},
      year={2025},
      eprint={2508.10076},
      archivePrefix={arXiv},
      primaryClass={cs.MS},
      url={https://arxiv.org/abs/2508.10076}, 
}

@article{Hastings,
  title = {Solving gapped Hamiltonians locally},
  author = {Hastings, M. B.},
  journal = {Phys. Rev. B},
  volume = {73},
  issue = {8},
  pages = {085115},
  numpages = {13},
  year = {2006},
  month = {Feb},
  publisher = {American Physical Society},
  doi = {10.1103/PhysRevB.73.085115},
  url = {https://link.aps.org/doi/10.1103/PhysRevB.73.085115}
}

@article{Kliesch,
  title = {Locality of Temperature},
  author = {Kliesch, M. and Gogolin, C. and Kastoryano, M. J. and Riera, A. and Eisert, J.},
  journal = {Phys. Rev. X},
  volume = {4},
  issue = {3},
  pages = {031019},
  numpages = {19},
  year = {2014},
  month = {Jul},
  publisher = {American Physical Society},
  doi = {10.1103/PhysRevX.4.031019},
  url = {https://link.aps.org/doi/10.1103/PhysRevX.4.031019}
}

@article{Molnar,
  title = {Approximating Gibbs states of local Hamiltonians efficiently with projected entangled pair states},
  author = {Molnar, Andras and Schuch, Norbert and Verstraete, Frank and Cirac, J. Ignacio},
  journal = {Phys. Rev. B},
  volume = {91},
  issue = {4},
  pages = {045138},
  numpages = {11},
  year = {2015},
  month = {Jan},
  publisher = {American Physical Society},
  doi = {10.1103/PhysRevB.91.045138},
  url = {https://link.aps.org/doi/10.1103/PhysRevB.91.045138}
}

@ARTICLE{Sandvik,
       author = {{Sandvik}, Anders W.},
        title = "{Stochastic Series Expansion Methods}",
      journal = {arXiv e-prints},
         year = 2019,
        month = sep,
          eid = {arXiv:1909.10591},
        pages = {arXiv:1909.10591},
          doi = {10.48550/arXiv.1909.10591},
archivePrefix = {arXiv},
       eprint = {1909.10591},
 primaryClass = {cond-mat.str-el},
       adsurl = {https://ui.adsabs.harvard.edu/abs/2019arXiv190910591S}
}

@article{Chen,
  title = {Exponential Thermal Tensor Network Approach for Quantum Lattice Models},
  author = {Chen, Bin-Bin and Chen, Lei and Chen, Ziyu and Li, Wei and Weichselbaum, Andreas},
  journal = {Phys. Rev. X},
  volume = {8},
  issue = {3},
  pages = {031082},
  numpages = {29},
  year = {2018},
  month = {Sep},
  publisher = {American Physical Society},
  doi = {10.1103/PhysRevX.8.031082},
  url = {https://link.aps.org/doi/10.1103/PhysRevX.8.031082}
}

@article{Czarnik_5,
  title = {Variational tensor network renormalization in imaginary time: Two-dimensional quantum compass model at finite temperature},
  author = {Czarnik, Piotr and Dziarmaga, Jacek and Ole\ifmmode \acute{s}\else \'{s}\fi{}, Andrzej M.},
  journal = {Phys. Rev. B},
  volume = {93},
  issue = {18},
  pages = {184410},
  numpages = {11},
  year = {2016},
  month = {May},
  publisher = {American Physical Society},
  doi = {10.1103/PhysRevB.93.184410},
  url = {https://link.aps.org/doi/10.1103/PhysRevB.93.184410}
}

@article{Czarnik_6,
  title = {Overcoming the sign problem at finite temperature: Quantum tensor network for the orbital ${e}_{g}$ model on an infinite square lattice},
  author = {Czarnik, Piotr and Dziarmaga, Jacek and Ole\ifmmode \acute{s}\else \'{s}\fi{}, Andrzej M.},
  journal = {Phys. Rev. B},
  volume = {96},
  issue = {1},
  pages = {014420},
  numpages = {8},
  year = {2017},
  month = {Jul},
  publisher = {American Physical Society},
  doi = {10.1103/PhysRevB.96.014420},
  url = {https://link.aps.org/doi/10.1103/PhysRevB.96.014420}
}

@article{Czarnik_8,
  title = {Time evolution of an infinite projected entangled pair state: An algorithm from first principles},
  author = {Czarnik, Piotr and Dziarmaga, Jacek},
  journal = {Phys. Rev. B},
  volume = {98},
  issue = {4},
  pages = {045110},
  numpages = {9},
  year = {2018},
  month = {Jul},
  publisher = {American Physical Society},
  doi = {10.1103/PhysRevB.98.045110},
  url = {https://link.aps.org/doi/10.1103/PhysRevB.98.045110}
}

@article{Czarnik_9,
  title = {Time evolution of an infinite projected entangled pair state: An efficient algorithm},
  author = {Czarnik, Piotr and Dziarmaga, Jacek and Corboz, Philippe},
  journal = {Phys. Rev. B},
  volume = {99},
  issue = {3},
  pages = {035115},
  numpages = {12},
  year = {2019},
  month = {Jan},
  publisher = {American Physical Society},
  doi = {10.1103/PhysRevB.99.035115},
  url = {https://link.aps.org/doi/10.1103/PhysRevB.99.035115}
}

@article{Czarnik_10,
  title = {Tensor network simulation of the Kitaev-Heisenberg model at finite temperature},
  author = {Czarnik, Piotr and Francuz, Anna and Dziarmaga, Jacek},
  journal = {Phys. Rev. B},
  volume = {100},
  issue = {16},
  pages = {165147},
  numpages = {9},
  year = {2019},
  month = {Oct},
  publisher = {American Physical Society},
  doi = {10.1103/PhysRevB.100.165147},
  url = {https://link.aps.org/doi/10.1103/PhysRevB.100.165147}
}

@article{Dziarmaga_NTU,
  title = {Time evolution of an infinite projected entangled pair state: Neighborhood tensor update},
  author = {Dziarmaga, Jacek},
  journal = {Phys. Rev. B},
  volume = {104},
  issue = {9},
  pages = {094411},
  numpages = {10},
  year = {2021},
  month = {Sep},
  publisher = {American Physical Society},
  doi = {10.1103/PhysRevB.104.094411},
  url = {https://link.aps.org/doi/10.1103/PhysRevB.104.094411}
}

@article{Dziarmaga_gradient,
  title = {Time evolution of an infinite projected entangled pair state: A gradient tensor update in the tangent space},
  author = {Dziarmaga, Jacek},
  journal = {Phys. Rev. B},
  volume = {106},
  issue = {1},
  pages = {014304},
  numpages = {8},
  year = {2022},
  month = {Jul},
  publisher = {American Physical Society},
  doi = {10.1103/PhysRevB.106.014304},
  url = {https://link.aps.org/doi/10.1103/PhysRevB.106.014304}
}

@article{Zhang,
  title = {Finite temperature dopant-induced spin reorganization explored via tensor networks in the two-dimensional $t\text{\ensuremath{-}}J$ model},
  author = {Zhang, Yintai and Sinha, Aritra and Rams, Marek M. and Dziarmaga, Jacek},
  journal = {Phys. Rev. B},
  volume = {113},
  issue = {8},
  pages = {085113},
  numpages = {13},
  year = {2026},
  month = {Feb},
  publisher = {American Physical Society},
  doi = {10.1103/6pcg-qq4p},
  url = {https://link.aps.org/doi/10.1103/6pcg-qq4p}
}

@ARTICLE{Wang_Cluster,
       author = {{Wang}, Ling and {Verstraete}, Frank},
        title = "{Cluster update for tensor network states}",
      journal = {arXiv e-prints},
         year = 2011,
        month = oct,
          eid = {arXiv:1110.4362},
        pages = {arXiv:1110.4362},
          doi = {10.48550/arXiv.1110.4362},
archivePrefix = {arXiv},
       eprint = {1110.4362},
 primaryClass = {cond-mat.str-el},
       adsurl = {https://ui.adsabs.harvard.edu/abs/2011arXiv1110.4362W}
}

@Article{Hubig_time,
	title={{Evaluation of time-dependent correlators after a local quench in iPEPS: hole motion in the t-J model}},
	author={Claudius Hubig and Annabelle Bohrdt and Michael Knap and Fabian Grusdt and J. Ignacio Cirac},
	journal={SciPost Phys.},
	volume={8},
	pages={021},
	year={2020},
	publisher={SciPost},
	doi={10.21468/SciPostPhys.8.2.021},
	url={https://scipost.org/10.21468/SciPostPhys.8.2.021},
}

@article{Kshetrimayum_time,
  title = {Time evolution of many-body localized systems in two spatial dimensions},
  author = {Kshetrimayum, A. and Goihl, M. and Eisert, J.},
  journal = {Phys. Rev. B},
  volume = {102},
  issue = {23},
  pages = {235132},
  numpages = {12},
  year = {2020},
  month = {Dec},
  publisher = {American Physical Society},
  doi = {10.1103/PhysRevB.102.235132},
  url = {https://link.aps.org/doi/10.1103/PhysRevB.102.235132}
}

@article{Kshetrimayum_thermal,
  title = {Tensor Network Annealing Algorithm for Two-Dimensional Thermal States},
  author = {Kshetrimayum, A. and Rizzi, M. and Eisert, J. and Or\'us, R.},
  journal = {Phys. Rev. Lett.},
  volume = {122},
  issue = {7},
  pages = {070502},
  numpages = {6},
  year = {2019},
  month = {Feb},
  publisher = {American Physical Society},
  doi = {10.1103/PhysRevLett.122.070502},
  url = {https://link.aps.org/doi/10.1103/PhysRevLett.122.070502}
}

@article{beliefpropagation,
  title = {Tensor networks contraction and the belief propagation algorithm},
  author = {Alkabetz, R. and Arad, I.},
  journal = {Phys. Rev. Res.},
  volume = {3},
  issue = {2},
  pages = {023073},
  numpages = {12},
  year = {2021},
  month = {Apr},
  publisher = {American Physical Society},
  doi = {10.1103/PhysRevResearch.3.023073},
  url = {https://link.aps.org/doi/10.1103/PhysRevResearch.3.023073}
}

@article{tindall,
  title = {Efficient Tensor Network Simulation of IBM's Eagle Kicked Ising Experiment},
  author = {Tindall, Joseph and Fishman, Matthew and Stoudenmire, E. Miles and Sels, Dries},
  journal = {PRX Quantum},
  volume = {5},
  issue = {1},
  pages = {010308},
  numpages = {16},
  year = {2024},
  month = {Jan},
  publisher = {American Physical Society},
  doi = {10.1103/PRXQuantum.5.010308},
  url = {https://link.aps.org/doi/10.1103/PRXQuantum.5.010308}
}

@ARTICLE{Naravane,
       author = {{Naravane}, Adwait and {Sugimoto}, Yuto and {Akiyama}, Shinichiro and {Haegeman}, Jutho and {Ueda}, Atsushi},
        title = "{Deconfinement from Thermal Tensor Networks: Universal CFT signature in (2+1)-dimensional $\mathbb{Z}_N$ lattice gauge theory}",
      journal = {arXiv e-prints},
         year = 2026,
        month = feb,
          eid = {arXiv:2602.13124},
        pages = {arXiv:2602.13124},
          doi = {10.48550/arXiv.2602.13124},
archivePrefix = {arXiv},
       eprint = {2602.13124},
 primaryClass = {hep-th},
       adsurl = {https://ui.adsabs.harvard.edu/abs/2026arXiv260213124N}
}

@ARTICLE{Zhang_preconditioner,
       author = {{Zhang}, Xing-Yu and {Yang}, Qi and {Corboz}, Philippe and {Haegeman}, Jutho and {Tang}, Wei},
        title = "{Accelerating two-dimensional tensor network optimization by preconditioning}",
      journal = {arXiv e-prints},
         year = 2025,
        month = nov,
          eid = {arXiv:2511.09546},
        pages = {arXiv:2511.09546},
          doi = {10.48550/arXiv.2511.09546},
archivePrefix = {arXiv},
       eprint = {2511.09546},
 primaryClass = {cond-mat.str-el},
       adsurl = {https://ui.adsabs.harvard.edu/abs/2025arXiv251109546Z}
}

@article{Frustration_Vanhecke_Colbois,
  title = {Solving frustrated Ising models using tensor networks},
  author = {Vanhecke, Bram and Colbois, Jeanne and Vanderstraeten, Laurens and Verstraete, Frank and Mila, Fr\'ed\'eric},
  journal = {Phys. Rev. Res.},
  volume = {3},
  issue = {1},
  pages = {013041},
  numpages = {13},
  year = {2021},
  month = {Jan},
  publisher = {American Physical Society},
  doi = {10.1103/PhysRevResearch.3.013041},
  url = {https://link.aps.org/doi/10.1103/PhysRevResearch.3.013041}
}

@article{Higher_order_hybrid,
  title = {Higher-order hybrid Monte Carlo algorithms},
  author = {Creutz, Michael and Gocksch, Andreas},
  journal = {Phys. Rev. Lett.},
  volume = {63},
  issue = {1},
  pages = {9--12},
  numpages = {0},
  year = {1989},
  month = {Jul},
  publisher = {American Physical Society},
  doi = {10.1103/PhysRevLett.63.9},
  url = {https://link.aps.org/doi/10.1103/PhysRevLett.63.9}
}

@article{Sinha2025,
  author       = {Sinha, Aritra and Wietek, Alexander},
  title        = {Forestalled phase separation as the precursor to stripe order},
  journal      = {Nature Communications},
  year         = {2025},
  volume       = {16},
  number       = {1},
  pages        = {10807},
  doi          = {10.1038/s41467-025-66563-5},
  url          = {https://doi.org/10.1038/s41467-025-66563-5},
  issn         = {2041-1723}
}

@article{ttJ,
  title = {Phase Diagram, $d$-Wave Superconductivity, and Pseudogap of the $t\ensuremath{-}{t}^{\ensuremath{'}}\ensuremath{-}J$ Model at Finite Temperature},
  author = {Qu, Dai-Wei and Li, Qiaoyi and Gong, Shou-Shu and Qi, Yang and Li, Wei and Su, Gang},
  journal = {Phys. Rev. Lett.},
  volume = {133},
  issue = {25},
  pages = {256003},
  numpages = {8},
  year = {2024},
  month = {Dec},
  publisher = {American Physical Society},
  doi = {10.1103/PhysRevLett.133.256003},
  url = {https://link.aps.org/doi/10.1103/PhysRevLett.133.256003}
}

@misc{PEPSKit,
      title={PEPSKit.jl: A Julia package for projected entangled-pair state simulations}, 
      author={Paul Brehmer and Lander Burgelman and Zheng-Yuan Yue and Gleb Fedorovich and Jutho Haegeman and Lukas Devos},
      year={2026},
      eprint={2605.19960},
      archivePrefix={arXiv},
      primaryClass={cond-mat.str-el},
      url={https://arxiv.org/abs/2605.19960}, 
}

@misc{dunham2025,
      title={Efficient Time Evolution of 2D Open-Quantum Lattice Models with Long-Range Interactions using Tensor Networks}, 
      author={Jack Dunham and Marzena H. Szymańska},
      year={2025},
      eprint={2512.01781},
      archivePrefix={arXiv},
      primaryClass={quant-ph},
      url={https://arxiv.org/abs/2512.01781}, 
}

@article{Sotnikov,
  title = {Single-layer tensor network approach for three-dimensional quantum systems},
  author = {Lukin, I. V. and Sotnikov, A. G.},
  journal = {Phys. Rev. B},
  volume = {110},
  issue = {6},
  pages = {064422},
  numpages = {11},
  year = {2024},
  month = {Aug},
  publisher = {American Physical Society},
  doi = {10.1103/PhysRevB.110.064422},
  url = {https://link.aps.org/doi/10.1103/PhysRevB.110.064422}
}

@article{Pollman2015,
  title = {Time-evolving a matrix product state with long-ranged interactions},
  author = {Zaletel, Michael P. and Mong, Roger S. K. and Karrasch, Christoph and Moore, Joel E. and Pollmann, Frank},
  journal = {Phys. Rev. B},
  volume = {91},
  issue = {16},
  pages = {165112},
  numpages = {8},
  year = {2015},
  month = {Apr},
  publisher = {American Physical Society},
  doi = {10.1103/PhysRevB.91.165112},
  url = {https://link.aps.org/doi/10.1103/PhysRevB.91.165112}
}

\appendix

\section{Details on the cluster expansion construction}
\label{ap:CE}

\subsection{Construction in 1D}

In order to make the construction of \ref{sec:methodology} more explicit, we will show the construction in 1D for $P = 3$. The construction in 2D is similar. We assume a nearest-neighbour Hamiltonian with no onsite interaction, i.e. the Hamiltonian of \eqref{eq:nn_hamiltonian}. The physical (virtual) legs will be drawn in the vertical (horizontal) direction. Since there is no onsite interaction, the restriction of the Hamiltonian to one site is the zero operator, and the $p = 1$ term is equal to $\mathds{1} = \exp(\mathds{O})$. The tensor that will be used in the finite-state machine, will thus be given by

\begin{equation}
    \vcenter{\hbox{\includegraphics[page=22, scale=1.3]{Typst/CE_1D_explanation.pdf}}} = \vcenter{\hbox{\includegraphics[page=23, scale=1.3]{Typst/CE_1D_explanation.pdf}}} \; ,
\end{equation}

where trivial legs of dimension 1 are added, which are indicated by the dotted lines.

To get the terms for $p = 2$, we need to consider the exponential of the Hamiltonian restricted to two sites, which contains only one nearest-neighbour term.
The equivalent of \eqref{eq:Construction_CE_H2} for a one-dimensional network is thus given by 

\begin{equation}
\begin{split}
    & \vcenter{\hbox{\includegraphics[page=1, scale=1]{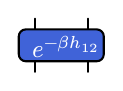}}} - \vcenter{\hbox{\includegraphics[page=2, scale=1]{Typst/CE_1D_explanation.pdf}}} \\ 
    &= \vcenter{\hbox{\includegraphics[page=3, scale=1]{Typst/CE_1D_explanation.pdf}}} = \vcenter{\hbox{\includegraphics[page=4, scale=1]{Typst/CE_1D_explanation.pdf}}}\; . \\
\end{split}
\label{eq:Construction_1D_CE_H2}
\end{equation}

We thus take the exponential for two sites and subtract the previous levels (which only contains $T_{00}$ for now). The result is a tensor $M^{(2)}$. Using the SVD, this yields two tensors via

\begin{equation}
    M^{(2)} = U S V^\dagger = \left( U \sqrt{S}\right) \left( \sqrt{S} V^\dagger \right) = A B
\end{equation}

Since $M^{(2)}$ is a $d^2 \times d^2$ tensor, its SVD yields $d^2$ singular values. Adding a trivial leg of dimension 1 on the left (right) side of tensor A (B) yields the tensor $T_{01}$ ($T_{10})$, i.e.

\begin{equation}
\begin{split}
    \vcenter{\hbox{\includegraphics[page=5, scale=1.3]{Typst/CE_1D_explanation.pdf}}} &\xrightarrow{} \vcenter{\hbox{\includegraphics[page=6, scale=1.3]{Typst/CE_1D_explanation.pdf}}} \; ,\\ 
    \vcenter{\hbox{\includegraphics[page=7, scale=1.3]{Typst/CE_1D_explanation.pdf}}} &\xrightarrow{} \vcenter{\hbox{\includegraphics[page=8, scale=1.3]{Typst/CE_1D_explanation.pdf}}} \; .\\
\end{split}
\end{equation}

To obtain the equivalent of \eqref{eq:Construction_CE_H3}, we take the exponential of the Hamiltonian restricted to three sites, thus containing a nearest-neighbour term acting on sites $1$ and $2$, and another term acting on sites $2$ and $3$. The terms that need to be subtracted are now all the combinations that can be made with tensors obtained up until now, i.e. $T_{00}$, $T_{01}$, and $T_{10}$. 

\begin{equation}
\begin{split}
    & \vcenter{\hbox{\includegraphics[page=10, scale=1]{Typst/CE_1D_explanation.pdf}}} - \vcenter{\hbox{\includegraphics[page=18, scale=1]{Typst/CE_1D_explanation.pdf}}} \\ 
    & - \vcenter{\hbox{\includegraphics[page=19, scale=1]{Typst/CE_1D_explanation.pdf}}} - \vcenter{\hbox{\includegraphics[page=20, scale=1]{Typst/CE_1D_explanation.pdf}}}\\
    &= \vcenter{\hbox{\includegraphics[page=14, scale=1]{Typst/CE_1D_explanation.pdf}}} = \vcenter{\hbox{\includegraphics[page=15, scale=1]{Typst/CE_1D_explanation.pdf}}} \; .\\
\end{split}
\label{eq:Construction_1D_CE_H3}
\end{equation}

The last line is a linear equation for $X$ and its solution will be called $T_{11}$. A more general form was given in \eqref{eq:Construction_CE_H3}, where an SVD of the matrix $M^{(2)}$ yields three different tensors. This leads to a different construction of the cluster expansion, which can be shown to be suboptimal \cite{ClusterExpansion2}.

The resulting Matrix Product Operator (MPO) representing the exponential of the Hamiltonian is given by a matrix containing the tensors $T_{00}$, $T_{01}$, $T_{10}$, and $T_{11}$, namely

\begin{equation}
    T =\left[
\begin{array}{c|c}
T_{00} & T_{01} \\ \hline
T_{10} & T_{11}
\end{array}\right] \; .
\label{eq:MPO_Matrix}
\end{equation}

This MPO reproduces the exponential exactly on three sites, which can easily be seen by contracting it over three sites and separating it from the rest of the infinite lattice by forcing the outer levels to be 0.

\begin{equation}
\begin{split}
    &\vcenter{\hbox{\includegraphics[page=16, scale=1]{Typst/CE_1D_explanation.pdf}}} = \sum_{k,l = 0}^1 \vcenter{\hbox{\includegraphics[page=17, scale=1]{Typst/CE_1D_explanation.pdf}}} \\
    &= \vcenter{\hbox{\includegraphics[page=18, scale=1]{Typst/CE_1D_explanation.pdf}}} + \vcenter{\hbox{\includegraphics[page=19, scale=1]{Typst/CE_1D_explanation.pdf}}} \\
    &+ \vcenter{\hbox{\includegraphics[page=20, scale=1]{Typst/CE_1D_explanation.pdf}}} + \vcenter{\hbox{\includegraphics[page=21, scale=1]{Typst/CE_1D_explanation.pdf}}} \\
    &= \vcenter{\hbox{\includegraphics[page=10, scale=1]{Typst/CE_1D_explanation.pdf}}} \; ,\\
\end{split}
\end{equation}

in which the last expression is equivalent to \eqref{eq:Construction_1D_CE_H3}. Forcing the outer levels to be 0 is done by contracting the network with the tensor

\begin{equation}
    \vcenter{\hbox{\includegraphics[page=24, scale=1]{Typst/CE_1D_explanation.pdf}}} =\left[
    \begin{array}{c}
    \mathds{1} \\ \hline
    \mathds{O} \\
    \end{array}\right] \; .
\end{equation}

\subsection{Dimension of the MPO}

Since the virtual Hilbert space of the MPO is given by the direct sum of the Hilbert spaces of the levels, i.e.

\begin{equation}
    \mathcal{H} = \bigoplus_{\ell = 0}^{\ell_{\text{max}}} \mathcal{H}_\ell \; ,
\end{equation}

its bond dimension is the sum of the dimensions of these Hilbert spaces. These are determined by the amount of non-zero singular values. As already mentioned, $\text{dim}(\mathcal{H}_1) \leq d^2$, with the inequality being saturated for generic hamiltonians. It can easily be seen that more generally, it holds that 

\begin{equation}
    \text{dim}(\mathcal{H}_{\ell}) \leq d^{2 \ell} \; ,
\end{equation}

from which it can also be seen that the dimension of the (trivial) virtual leg of level 0 is equal to $\text{dim}(\mathcal{H}_{0}) = 1$. It can thus be concluded that 

\begin{equation}
    D_2 \leq \sum_{\ell = 0}^{\ell_{\text{max}}} d^{2 \ell} \; .
\end{equation}

The highest level that takes part in a cluster expansion of order $P$ is given by $\ell_{\text{max}} = \lfloor \frac{P}{2} \rfloor$. Since the physical dimensions of the spinless fermion model, the tJ model, and the Hubbard model are 2, 3, and 4, this equation leads to a bond dimension of 5, 10, and 17, respectively, for a $P = 3$ cluster expansion.

\vspace{5mm}
\section{Finite entanglement scaling}
\label{ap:FCLS}

Given an accurate representation of the thermal density operator $\rho(\beta)$ with bond dimension $D$, an accurate environment should be found in order to calculate observables. If the boundary MPS is calculated using VUMPS, choosing a finite environment bond dimension $\chi$ for this MPS $\ket{\psi_B(A)}$ will introduce errors in the observables. These errors could be reduced by plotting the observables as a function of an error measure, postulating a scaling hypothesis, and extrapolating the results towards error measure zero. While a possible error measure is the inverse environment bond dimension $\frac{1}{\chi}$, a more natural choice can be found by considering the transfer matrix $T = \sum_{s=1}^d A^s \otimes \bar{A}^s$. If we denote the eigenvalues of this transfer matrix, sorted in decreasing magnitude, as 
\begin{equation}
    \lambda_j = e^{-\epsilon_j + i \phi_j}, 
\end{equation}
for $j = 0, \ldots, \chi^2-1$, where $\lambda_0 = 1$ for a properly normalized MPS, then we can then define
\begin{eqnarray}
    \xi &= 1 / \epsilon_1 \; ,\\ 
    \delta &= \epsilon_2 - \epsilon_1 \; ,
\end{eqnarray}
where $\xi$ is the correlation length and $\delta$ is related to the maximal entanglement the simulation can capture \cite{Rams_2018}, with ${\lim}_{D,\chi \to \infty} \delta = 0$. The $\chi\to\infty$ extrapolation can thus be recast as a $\delta\to 0$ extrapolation, with $1/\delta$ behaving more naturally as a length scale. 

\end{document}